\documentclass[preprint]{aastex}

\newcommand{\kms}{km~s$^{-1}$}

\newcommand{\teff}{$T_{\rm{eff}}$}
\newcommand{\grav}{log($g$)}
\newcommand{\etal}{et al.}
\newcommand{\eqw}{$W_{\lambda}$}

\newcommand{\mystar}{OGLE--2007--BLG--349S}
\newcommand{\johnstar}{OGLE--2006--BLG--265S}
\newcommand{\moa}{MOA--2006--BLG--099S}
\newcommand{\fehogle}{$+0.51\pm 0.09$}

\begin{document}

\title{Clues to the Metallicity Distribution in the
Galactic Bulge: Abundances in \mystar
\altaffilmark{1}}

\author{Judith G. Cohen\altaffilmark{2}, 
Wenjin Huang\altaffilmark{2}, A. Udalski\altaffilmark{4},
Andrew Gould\altaffilmark{3} \& Jennifer A. Johnson\altaffilmark{3}  }

\altaffiltext{1}{Based in part on observations obtained at the
W.M. Keck Observatory, which is operated jointly by the California 
Institute of Technology, the University of California, and the
National Aeronautics and Space Administration.}

\altaffiltext{2}{Palomar Observatory, Mail Stop 105-24,
California Institute of Technology, Pasadena, Ca., 91125, 
jlc,wenjin@astro.caltech.edu}

\altaffiltext{3}{Department of Astronomy, Ohio State
University, 140 W. 18the Ave., Columbus, OH 43210; 
gould,jaj@astronomy.ohio-state.edu}

\altaffiltext{4}{Warsaw University Observatory, A1. Ujazdowskie 4,
00-478 Waszawa, Poland; udalski@astrouw.edu.pl,}

\begin{abstract}
We present an abundance analysis based on high dispersion and high
signal-to-noise ratio Keck spectra of a very highly microlensed Galactic
bulge dwarf, \mystar, with
\teff $\sim 5400$~K.  The amplification at the time
the spectra were taken ranged from 350 to 450.
This bulge  star is highly enhanced
in metallicity with [Fe/H]\footnote{We adopt the usual spectroscopic notations
that [A/B] ~ $\equiv ~ log_{10} (N_A/N_B)_* - log_{10} (N_A/N_B)_{\odot}$, 
and that log$[\epsilon(A)] ~  \equiv ~ log_{10} (N_A/N_H) + 12.00$, for elements
$A$ and $B$.}  = \fehogle ~dex.  The abundance ratios for the
28 species of 26 elements for which features could be detected in
the spectra are almost all solar.  In particular, there is no evidence 
for  enhancement of any of the $\alpha$-elements including O and Mg.   
We conclude   that the high [Fe/H] seen in this star, when 
combined with the equally high [Fe/H] derived in 
previous detailed abundance analysis of two other
Galactic bulge dwarfs, both also highly magnified by microlensing,
implies that
the median metallicity in the Galactic bulge is very high.
We thus infer that
many previous estimates of the metallicity  distribution
in the Galactic bulge have substantially underestimated the mean 
Fe-metallicity there
due to sample bias, and suggest a  candidate mechanism for such.
If our conjecture proves valid, it may be necessary to update the 
calibrations for the algorithms  
used by many groups to interpret   spectra and 
broad band photometry of the integrated light
of very metal-rich old  stellar populations, including
luminous elliptical galaxies.

\end{abstract}


\section{Introduction}

Microlensing occurs when a ``lens'' (star, planet, black hole, etc)
becomes closely aligned with a more distant ``source'' star, whose image
it both magnifies and distorts.  Normally the observer is most interested
to learn about the lens, but microlensing data can simultaenously serve as
a powerful probe of the source.  If the source transits the lens
(or the ``caustics'' generated by the lens, in the case of binary lenses), 
then it resolves the source, allowing detailed
limb-darkening profiles from a photometric times series 
\citep{fields03}
or even spectral resolution of surface features like the chromosphere
\citep{cassan04} 
from judiciously taken spectra.

The lens can
also serve simply to amplify the role of the telescope as a ``light bucket''.
\citet{minniti98} 
provocatively titled their report on observations
of a bulge dwarf that was magnified by a factor $A=2.25$ 
``Using Keck I as a 15m Diameter Telescope'' and \cite{cavallo}
presented a preliminary abundance analysis for six stars
with small magnification ($2.5 < A < 30$), including this
star and two other bulge dwarfs.  
Obviously, this trick
could in principle be improved to arbitrary ``diameters'' simply by
observing the events at higher magnification.  The problem is that it
is extremely difficult to recognize high-magnification events in advance,
and harder still to activate large-telescope observations in time to
take advantage of them.  The process has been facilitated by microlensing
planet hunters, who prize high-magnification events because of their
extreme sensitivity to planets 
\citep{ob05071,ob05169,ob06109}.  \citet{johnson07}
were the first to piggy-back on the microlensing planet hunters, obtaining
a 15 minute Keck spectrum of the Galactic bulge dwarf \johnstar\
at magnification $A=135$.  This star, the first bulge dwarf with a 
high-quality spectrum, proved to be extremely metal-rich, a fact that 
might be telling us that the bulge is much more metal-rich than seems
indicated by available spectra of giants, but could also just
be the ``luck of the
draw''.  The 
\citet{johnson07} 
results therefore substantially raised the
premium on obtaining highly-magnified spectra of bulge dwarfs.

On 2 July 2007, the OGLE 
collaboration\footnote{http://www.astrouw.edu.pl/$\sim$ogle/ogle3/ews/ews.html}
\citep{ews}
announced OGLE--2007--BLG--349\ (RA=18:05:24.43; DEC = $-26$:25:19.0)
at Galactic coordinates $(l,b)=(4.4,-2.5)$, i.e., 5.1$^{\circ}$
from the Galactic center, as
a probable microlensing event.  The Microlensing Follow Up 
Network\footnote{http://www.astronomy.ohio-state.edu/$\sim$microfun/}
($\mu$FUN)
began monitoring the event on 18 August to determine whether the event
would be high-magnification and on 3 September issued a general alert that
it would reach at least $A>200$ two nights hence.  On this basis, $\mu$FUN
organized world-wide photometric observations, whose outcome will be reported
elsewhere \citep{dong08}
and also contacted JGC at the Keck telescope to recommend intensive 
observations.

The ability to obtain high resolution, high quality spectra
of Galactic bulge stars and to carry out a detailed abundance
analysis offers an unbiased way to determine the metallicity
distribution of stars in the Galactic bulge, as well as their
detailed chemical inventory.  Our abundance
analysis of \mystar\ is described  in the first few sections
of the paper,
with the key results presented in \S\ref{section_abund_results}.
In an effort to explain our rather surprising results, we indicate in
\S\ref{section_discussion} how past studies of the brightest giants, which are the only
bulge stars for which detailed abundance analyses can be derived from
spectra obtained under normal conditions, may be subject to previously
ignored selection effects, and how this might impact
studies of the integrated light for metal-rich old simple stellar populations.  
Abundance ratios in
the Galactic bulge giants and in the microlensed dwarfs
are discussed in \S\ref{section_abund_ratios}.
A brief summary concludes the paper, 
while an  appendix discusses the behavior of selected diffuse interstellar
bands in the spectrum of \mystar.

\section{Observations}

In light of the prediction that the magnification of the presumed
Galactic bulge star \mystar\ would be very large,
we decided to observe the object in an attempt to obtain
a high signal-to-noise and high resolution spectrum.
The night of 5 September 2007 was clear at the Keck Observatory, and the 
seeing\footnote{The seeing was measured from the profile of the
stellar image along the slit in the 
three individual spectroscopic exposures of \mystar.}
was good, 0.8 arcsec in the optical.
We found the microlensed star, confirmed that it was bright, and
took three consecutive spectra, each 1350 sec in length.  The UTCs at the end of
each exposure were 
06:03:14,  06:27:04 and 06:50:27. 
HIRES-R \citep{vogt94} was used on the Keck~I telescope in 
a configuration with coverage from 3900 to 8350~\AA, with
small gaps between the orders beyond 6650~\AA. 
The slit was 0.86~arcsec wide, giving a spectral resolution of
48,000.

The magnifications at the times of the
3 spectra were A=350, A=390, and A=450; apparent $I$ at these times
ranged from
from 13.68 to 13.41~mag \citep{dong08}. Hence our observations
were carried out at a time when the amplification was larger by
a factor of 3 
than that of any previous  highly magnified star with a high
resolution spectrum.  While the SNR of the resulting summed spectrum is low
for $\lambda < 5000$~\AA\
($\sim 30$/spectral resolution element at 4500~\AA), not unexpected given the
high extinction toward the Galactic bulge, the SNR/spectral resolution element
in the continuum at $\lambda > 5500$~\AA\
exceeds 90.  Isolated lines
with \eqw\ of 15~m\AA\ are easily detected at such wavelengths.

According to the microlensing model of \cite{dong08},
the source passed very close to a cusp shortly
(roughly 2 hrs) after the third observation.  This means
that the limb of the star
was magnified more strongly than the center.  
We have calculated the maximum effect
on the $V-I$ color due to differential magnification of
the (cooler) limb of the star being more magnified than the
center,  with limb-darkening included,  
to be less than 0.005~mag during the
spectroscopic exposures, and we ignore it.


\section{Stellar Parameters \label{section_params} }

We determined \teff\ for \mystar\ in three different ways,
from line depth
ratios of close pairs of temperature sensitive lines,
from Fe~I excitation, and from the ionization equilibrium
between Fe~I and Fe~II.

\cite{gray91} have demonstrated that ratios of the central
depth of selected pairs of lines that are close together in
wavelength and highly sensitive to temperature can be used to
deduce \teff\ to high accuracy for cool stars.  We checked
the 15 such line pairs studied by \cite{biazzo07}.
In many cases one or both of the lines were highly saturated
in the spectrum of \mystar,  with
central depths approaching 70\% of the continuum.  There were only 3 line
pairs for which both lines had central depths of 40\% or less.  
Using their polynomial fit for non-rotating dwarfs, we find a mean \teff\
from these three pairs of 5350~K.

The Fe abundance was determined from 135 lines of Fe~I 
and 11 of Fe~II.  As described in detail in \S\ref{section_abunds},
\teff = 5400~K produces a deduced Fe abundance
independent of the excitation potential of the Fe~I line
and a difference between [Fe/H] from Fe~I versus from Fe~II 
lines of only 0.08~dex, well within the expected uncertainty of
$\mid$ [Fe/H:Fe~I] -- [Fe/H:Fe~II] $\mid$.

These three methods give a mean \teff\ of 5431~K, with $\sigma = 72$~K.
We adopt \teff = 5400$\pm100$~K.
The extinction corrected $V-I$ color
can be determined from the light curve of the microlensed star.
The extinction to the source is taken to be that of the red clump
 stars on the horizontal branch in the neighborhood of the source, which is determined
 from OGLE~II publicly available multi-color photometry .  We assume that 
there is no differential extinction between the adjacent bulge 
red clump stars\footnote{The dereddened red clump in the Galactic bulge
is assumed to have $I_0 = 14.32$~mag and $(V-I)_0 = 1.00$~mag.}
and the target.  With this extinction,
the model of \cite{dong08} that fits the two-color ($V$ and $I$,
where the latter is in the Cousins system) microlensing light curve
for \mystar\  fixes the flux level
of the source to be $I_0 = 18.72$~mag with $(V-I)_0 = 0.73$~mag.  
Because the lens lies in the foreground disk  \citep{dong08},
the most probable distance to the source is that of the Galactic center,
8.0~kpc.
The absolute magnitude is then
$M_I = 4.22$.  The bulge star is thus slightly fainter and 
slightly redder than 
the Sun ($V-I = 0.688\pm 0.014$~ mag, \citealt{holmberg06}).
Combining this with the colors of stars along a metal-rich isochrone of
\cite{yi01} suggests \teff\ = 5550~K, 1.5$\sigma$ hotter than our adopted
\teff.  If the extinction corrected
$V-I$ color determined in this way is underestimated by 
by 0.08~mag 
(of which 0.05~mag  may arise from an underestimate
of the intrinsic color of the red clump)
as was suspected by \cite{johnson07}, who employed
exactly the same method for \johnstar, 
then the deduced \teff\ for \mystar\ would
be $\sim 5430$~K, essentially identical to that
deduced from the spectrum itself.

We next turn to the surface gravity. With \teff\ set, we can determine
\grav\ from the ionization equilibrium between Fe~I and Fe~II, from
an isochrone, and from the line profile in the damping wings
of a very strong line.   The Galactic bulge is old
\citep[see, for example,][]{ortolani95}.  We assume 
the age of \mystar\ is 9~Gyr.
From the isochrones of \cite{yi01}, we find that the main sequence
turnoff occurs at 5500~K for $Z = 0.04$ (a metallicity 2.4 times solar)
and no $\alpha$-enhancement
at an age of 9~Gyr.  
A star at 5400~K can be a subgiant with
\grav = 4.0 dex, or a dwarf below the turnoff
with \grav= 4.4 dex.  For an upper main sequence star with a fixed \teff,
\grav\ decreases by $\sim$0.15~dex when the isochrone age is changed
from 4 to 9~Gyr.  Similarly, \grav\ for a fixed age within this
interval and metallicity ranging from $z=0.00$ to $z=0.06$ increases by
$\sim$0.10~dex.  Thus our choice of an age of 9~Gyr for the isochrone
we adopt to determine \grav\ is not crucial, nor 
is the exact choice of metallicity
of the isochrone.

\begin{figure}
\epsscale{1.0}
\plotone{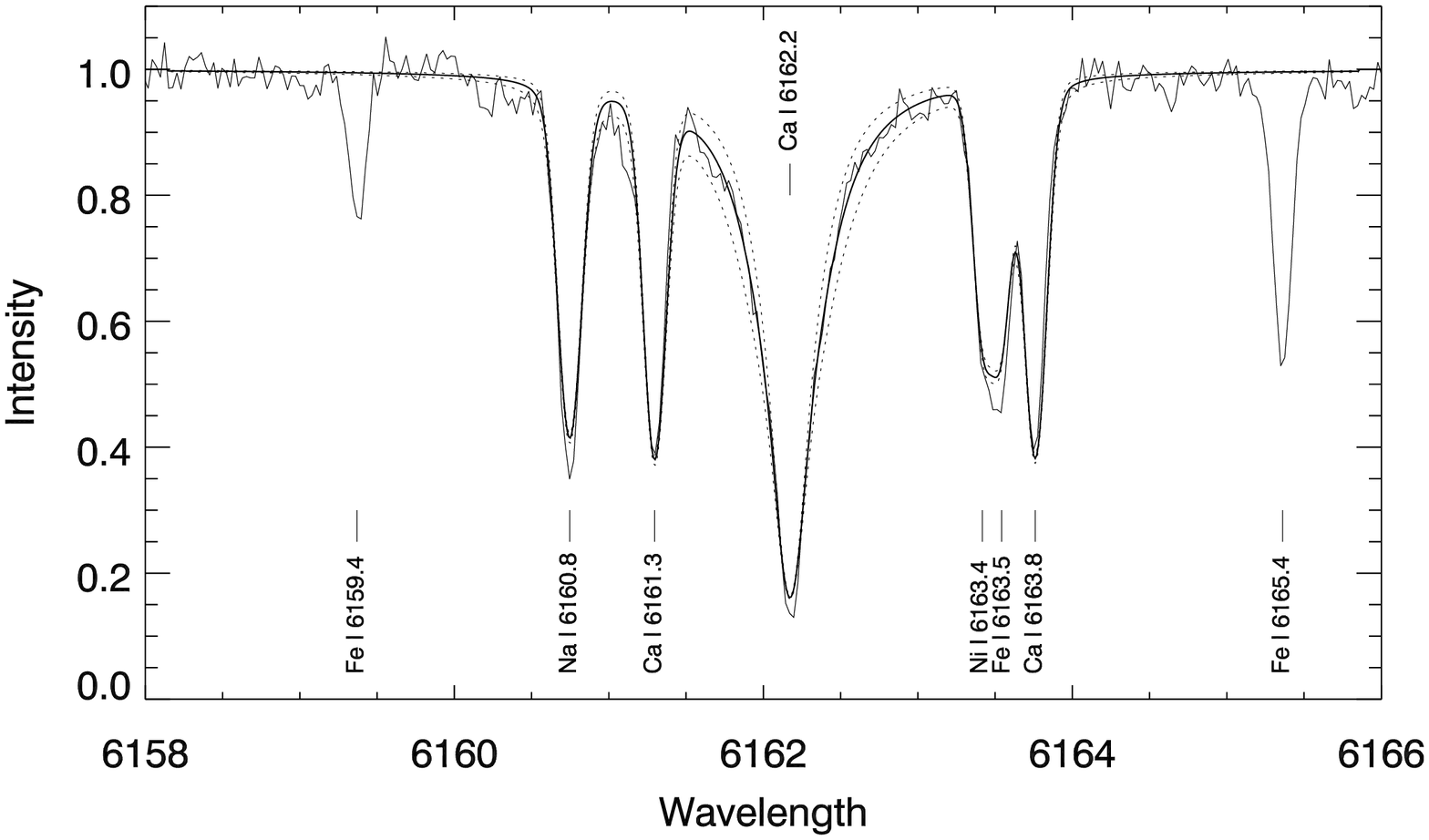}
\caption[]{The spectrum of \mystar\ in
the region of the strong Ca~I line at 6162.2~\AA\ is shown
shifted to the rest frame.  Overplotted
as a thick line is a
a spectral synthesis covering 6160 to 6165~\AA\
for \teff\ = 5400~K, [Fe/H] = +0.5~dex,
and solar [Ca/Fe] with the best fit \grav\ of 4.35~dex.
The thin dashed lines represent offsets
in \grav\ of $\pm0.3$~dex.
\label{figure_ca6160}}
\end{figure}

Profiles of very strong lines with obvious damping wings can
constrain \grav.  Figure~\ref{figure_ca6160} shows a spectral synthesis
in the region of the 6162.2~\AA\ line of Ca~I.  A synthesis
of the solar spectrum with the same line list gives a very good fit. 
The best fit for
\grav\ is 4.35~dex, supporting \mystar\ being a dwarf rather
than a subgiant.  Note that a solar [Ca/Fe] ratio and [Fe/H] = +0.5~dex
is assumed for this synthesis.
In addition, the
$M_I$ of 4.22~mag deduced from the extinction combined with the microlensing
light curve is 0.1~mag brighter than that of a 5400~K dwarf with
this isochrone, but is 1.0~mag fainter than that of a subgiant
from that isochrone with that \teff.
The predicted number density for dwarfs as compared to subgiants is
about 4:1 for a Salpeter IMF, further favoring the star being a dwarf.
We thus feel confident that \mystar\ is a dwarf just below the main 
sequence turnoff 
with \grav $\sim 4.4$~dex. The ionization equilibrium
between Fe~I and Fe~II is satisfactory at \grav\ = 4.5~dex.  The uncertainty
in \teff\ of 100~K translates into an uncertainty in
\grav\ of 0.2~dex.   We adopt \grav\ = 4.5$\pm0.2$~dex for the
abundance analysis.


\section{Abundance Analysis \label{section_abunds}}

Our Keck/HIRES spectrum of \mystar\ is full of strong lines with considerable
blending and crowding, especially in the blue.
Fe~I  lines at $\lambda < 5900$~\AA\ were ignored, as were
Fe~II lines blueward of 5100~\AA.   For all other species,
lines blueward of 5200~\AA\ were excluded.  A line list for Fe~I
and Fe~II was assembled from that used by
\cite{fulbright06} in their study of K giants in Baade's window
 and from JGC's personal line list,
augmented by selected very weak  isolated lines found in the
solar spectrum.  Equivalent widths and atomic parameters for the lines
used are given in Table~\ref{table_eqw}.  Lines with
\eqw\ $> 180$~m\AA\ were rejected unless the species
has very few detected features; those retained that
are stronger than this are the only detected line of K~I and the 5680~\AA\ Na~I
doublet.  
(The NaD lines are much too strong to consider
using and are hopelessly contaminated with interstellar absorption.)
Equivalent
widths were measured using an automatic  Gaussian fitting routine
combined with the measured heliocentric  $v_r$ of +99.47$\pm0.05$~\kms.
The stronger lines were all checked by hand to make sure
the damping wings were picked up when appropriate. 
The Mg~I triplet lines in the region of 6320~\AA, where there
is a broad autoionization feature of Ca, were measured 
by hand.  For elements with only a few detected lines,
all features were checked by hand as well.  The FTS solar spectrum of
\cite{wallace98}, available online, was very useful for this
purpose. The major uncertainty
in the \eqw\ results from the definition of the continuum in
the crowded spectrum of \mystar.

The abundance analysis
was carried out differentially with respect to the Sun,
since the \teff\ difference is only $\sim$400~K, and both
stars are dwarfs.
We used a current version of the LTE
spectral synthesis program MOOG \citep{moog}.
We employ the grid of stellar atmospheres from \cite{kurucz93}
with [Fe/H] = +0.5~dex and solar
abundance ratios  
without convective overshoot \citep{no_over}  and with the most recent opacity
distribution functions.  For consistency,
the solar model was taken from the appropriate grid of
Kurucz as well.

Spectral syntheses were used for the S~I doublet near 6750~\AA\
and for the Rb~I resonance line near 7800~\AA, which lies in
the red wing of a much stronger Si~I line.  A synthesis
of the CH band near 4320~\AA\ using the molecular line list
of \cite{ch} was carried out to determine
the C abundances.

Transition probabilities from the NIST Atomic Spectra 
Database Version 3.1 \citep{nist}
were used in general when available.   Since we analyze \mystar\ differentially
with respect to the Sun, the adopted $gf$ values are not crucial,
but when ``accurate'' values are available, we  used them.  
Accurate $gf$ values could not be located for a small number of lines
for which we were forced to resort to an inverted solar analysis to assign such; these
cases are indicated in Table~\ref{table_eqw}.  We used
damping constants from \cite{barklem00} when available.

We set the microturbulent velocity $v_t$ to 1.0~\kms, following the example
of the Sun.  This gave deduced Fe abundances independent of \eqw\
to within the accuracy that the resulting set of log[$\epsilon$(Fe)] 
from each of Fe~I lines could be measured.
Hyperfine structure corrections were used for the lines of Ba~II, Co~I, Cu~I,
Mn~I, Rb~I, Sc~II, and V~I.  These were taken from the
compilation generated by \cite{prochaska}, except that the
isotopic and HFS splittings
for the 7800~\AA\ Rb~I line were taken from \cite{lambert76}.

Non-LTE corrections were not generally
included, as this is a differential analysis with 
respect to Sun, and the stellar parameters of \mystar\ are fairly close
to those of the Sun.  Two key elements for which this might be an issue
are the 7770~\AA\ O~I triplet and our use of the  resonance doublet
of K~I.  We have checked the case of  O using 
the fitting formula given in equation (2) of
\cite{bensby04}.  When
all the differences between the stellar parameters for the OGLE star and the
Sun are taken into account, they yield  a negligible difference 
in the non-LTE correction for
this O triplet.  Two sets of non-LTE calculations are available
for the K~I resonance doublet, of which we observed the redder line;
the bluer line is in the middle of a strong terrestrial absorption band.
The results of the two groups,
given in \cite{ivanova}  and in \cite{takeda01}, are consistent and 
suggest that for solar metallicity
the K abundance of the OGLE star with respect to the Sun 
from a LTE analysis should
be increased by about 0.08~dex to take into account the
probable difference in the non-LTE correction for this line.  This
has  been implemented here.  However this adopted value must
be regarded as  still uncertain since the non-LTE corrections
for K  are metallicity
sensitive, and none are available for super-solar metallicity.

\section{Results of the Abundance Analysis \label{section_abund_results} }

The  final deduced abundances for \mystar\ for 28 species of
26 elements are given in Table~\ref{table_abunds}.
Our derived absolute abundances, the abundances relative to the Sun, and 
the abundance ratios [X/Fe] are given in this table. 
The abundance ratios use either 
Fe~I or Fe~II as the reference depending on the ionization
state and mean excitation potential of the measured lines of species
under consideration.  The 1$\sigma$ dispersion
around the mean for each species is given as $\sigma_{obs}$.
This is calculated from the set of differences between the
deduced solar abundance for the species in question and that 
found for \mystar\ for each observed line of the species.  Thus
neither random nor systematic errors in the $gf$  values contribute
to $\sigma_{obs}$.

While the absolute abundance for a given species listed 
in Table~\ref{table_abunds} will be affected
by any systematic error in the $gf$ values of the lines we
use here,  relative abundances [X/Fe] will not
since we have carried out a differential analysis with
respect to the Sun. An uncertainty
for [X/Fe] for each species, $\sigma_{pred}$, is calculated 
summing five terms  combined in quadrature representing
a change in \teff\ of 100~K, the corresponding uncertainty in \grav\
of 0.2~dex, a change in $v_t$ of 0.2~\kms, and 
a potential 0.25~dex  mismatch between
[Fe/H] of \mystar\ versus the value +0.5~dex of the model atmospheres
we are using.  The fifth term, the contribution for errors in \eqw, is 
set to 0.05~dex if only one or two
lines were measured; for a larger number of
detected lines we adopt $\sigma_{obs}/\sqrt{N(lines)}$
for this term.  This is added in quadrature to the other
four terms.

Uncertainties in the absolute abundances have only been calculated
for Fe as inferred from Fe~I and from Fe~II lines and are given below.

The sensitivity
of the deduced abundances to changes in the stellar parameters
or in $v_t$ is shown in Table~\ref{table_sens_abs}
for absolute abundances and in Table~\ref{table_sens_rel} for
relative abundances [X/Fe].  As expected, the
latter show much weaker dependences on the choice of stellar
parameters etc.   These entries were used to
generate the values of $\sigma_{pred}$ in the last column
of Table~\ref{table_abunds}.  Table~\ref{table_sens_abs}
shows that
the Fe~I abundance log$[\epsilon(Fe)]$ is almost independent of the choice of 
\teff, increasing by only +0.12~dex when
\teff\ is increased by 250~K. The neutral species of elements with resonance 
or low-excitation lines (K~I, V~I, Rb~I, and Zr~I) show the largest
increase in deduced abundance when \teff\ is increased, while those
species with only high excitation lines (the near-IR O triplet, C~I, and S~I) show
the opposite dependence on \teff, as expected.

The ionization
equilibrium for Fe   for the adopted
\teff\ and \grav\ is good, as is the Fe~I excitation equilibrium.  

In addition to presenting what we believe to be
the first abundance determinations for Rb and for Zr
for any star in the Galactic bulge, there
are three key results of the detailed abundance analysis
we have carried out
of \mystar.  The first is the very high metallicity we derive.
We find  [Fe/H] is \fehogle ~dex from 135 Fe~I lines and $+0.43 \pm0.17$~dex
from 11 Fe~II lines. To achieve a reduction of [Fe/H] to $< +0.3$~dex,
\teff\ must be reduced to ${\sim}5000$~K, with a change of \grav\ of +0.2~dex for
the dwarf or $-$0.1~dex
for the subgiant, for the star to lie on the
appropriate isochrone.  The Fe ionization equilibrium will then be
altered by 0.3~dex for the dwarf case and by somewhat less for the subgiant.
This will produce a Fe ionization equilibrium that is so far from
equality that the altered set of stellar parameters must be rejected.
To achieve solar metallicity and reasonable Fe ionization
equilibrium is simply not possible with any set of stellar parameters
that lie on the relevant isochrone.

Second the $\alpha$-elements O and Mg do not show any
excess with respect to Fe; 
oxygen, which is the most abundant $\alpha$-element, has sub-solar
[O/Fe].  Finally,
in all species for which the abundance is regarded as well determined
(the notes in the last column of Table~\ref{table_abunds} indicate the major
concerns),
there are no  credible deviations
from the solar ratios.

We note that [Fe/H] cannot be affected by mixing
of nuclear processed material from the interior of a dwarf or a red giant
as none would have been produced up to that stage of stellar
evolution.
Furthermore, any diffusion of Fe into the more quiescent dwarf as compared
to the stirred up outer layers of the giant would
be small in such high metallicity stars and would only
make the present conundrum of the unexpectedly high [Fe/H] 
we find in \mystar\  worse.  We also note that the three
line ratios we used as one method for determining \teff\ 
detailed in \S\ref{section_params} each
consisted of a V~I line paired with either a  Fe~I, Ni~I or Si~I line.
We find [V/Fe] for \mystar\ to be +0.17$\pm0.14$~dex
(Table~\ref{table_abunds}). If this ratio 
is actually slightly
above solar,  a slight underestimate of \teff\ would result from the
line ratio method.  Underestimating \teff\ is equivalent to
underestimating [Fe/H], which again makes the discrepancy with the
bulge giants worse.

\section{\mystar\ and The Metallicity Distribution of the 
Galactic Bulge \label{section_discussion} }

An early attempt to determine the metallicity distribution
in the Galactic bulge was that of \cite{sadler96},
who used  low resolution spectroscopy for 268 bulge giants and red clump stars
to derive a mean [Fe/H] of $-0.11\pm0.04$~dex.  \cite{ramirez00}
studied a sample of M giants in the near-IR; they found
a very similar mean [Fe/H] of $-0.21$~dex with a dispersion
of 0.30~dex.
\cite{zocalli} used extensive optical and near-IR photometry with
CMD fitting;
they derived a somewhat lower mean [Fe/H].
In all these cases, the calibration of the metallicity scale
relied on Galactic globular
clusters. \cite{zocalli} suggest that the differences
between several of these studies depend crucially on the abundances
adopted for the two highest metallicity GCs with high resolution
abundance analyses, NGC~6553 and NGC~6558.  [Fe/H] values for
these two GCs have ranged over more than 0.5~dex
in the literature, but
since the work of \cite{cohen99} and \cite{carretta01},
who suggested values higher  than most
previous studies, more recent analyses have
settled toward the higher values, see, e.g., 
\cite{zoccali04} and \cite{vlt_6553}. 

An early high dispersion spectroscopic study of
Galactic bulge K giants is that of \cite{mcwilliam94}, 
who found a mean [Fe/H]
in their sample of $-0.25$~dex. 
\cite{fulbright06} updates and expands upon this
earlier work.  They then use their detailed
abundance analyses of 27 K giants in Baade's window
to recalibrate the metallicities for the
much larger samples of \cite{sadler96}
and of \cite{rich88}.  The mean [Fe/H] they thus
deduce is $-0.10\pm0.04$~dex.  The median [Fe/H] of their sample
is also sub-solar.  As infrared
echelle spectrographs have become available on 8-m class
telescopes, high dispersion studies of Galactic bulge giants in the near-IR
have become possible, see, e.g.,
\cite{rich05}, \cite{cunha06} and \cite{rich07}.  The
mean [Fe/H] from the  sample of M giants studied by 
\cite{rich07} in the Galactic bulge
is well below solar, with [Fe/H] $-0.22\pm0.01$~dex.  The median Fe-metallicity
of the small sample of giants with near-IR spectra analyzed by \cite{cunha06}
is  also slightly  below solar metallicity.  Since this sample
largely overlaps that of \cite{fulbright06} it
is not included in the figures.

Very recently, \cite{zoccali08} have presented initial results
of a survey of Fe-metallicity in the Galactic bulge
from spectra with $\lambda/\Delta\lambda = 20,000$ of about  800 stars.
They find a radial gradient in [Fe/H] within the bulge
with the mean value going from +0.03~dex at $b = -4^{\circ}$ to
$-0.12$~dex at $b = -6^{\circ}$, and a sharp cutoff towards higher
metallicities.

\begin{figure}
\epsscale{1.0}
\plotone{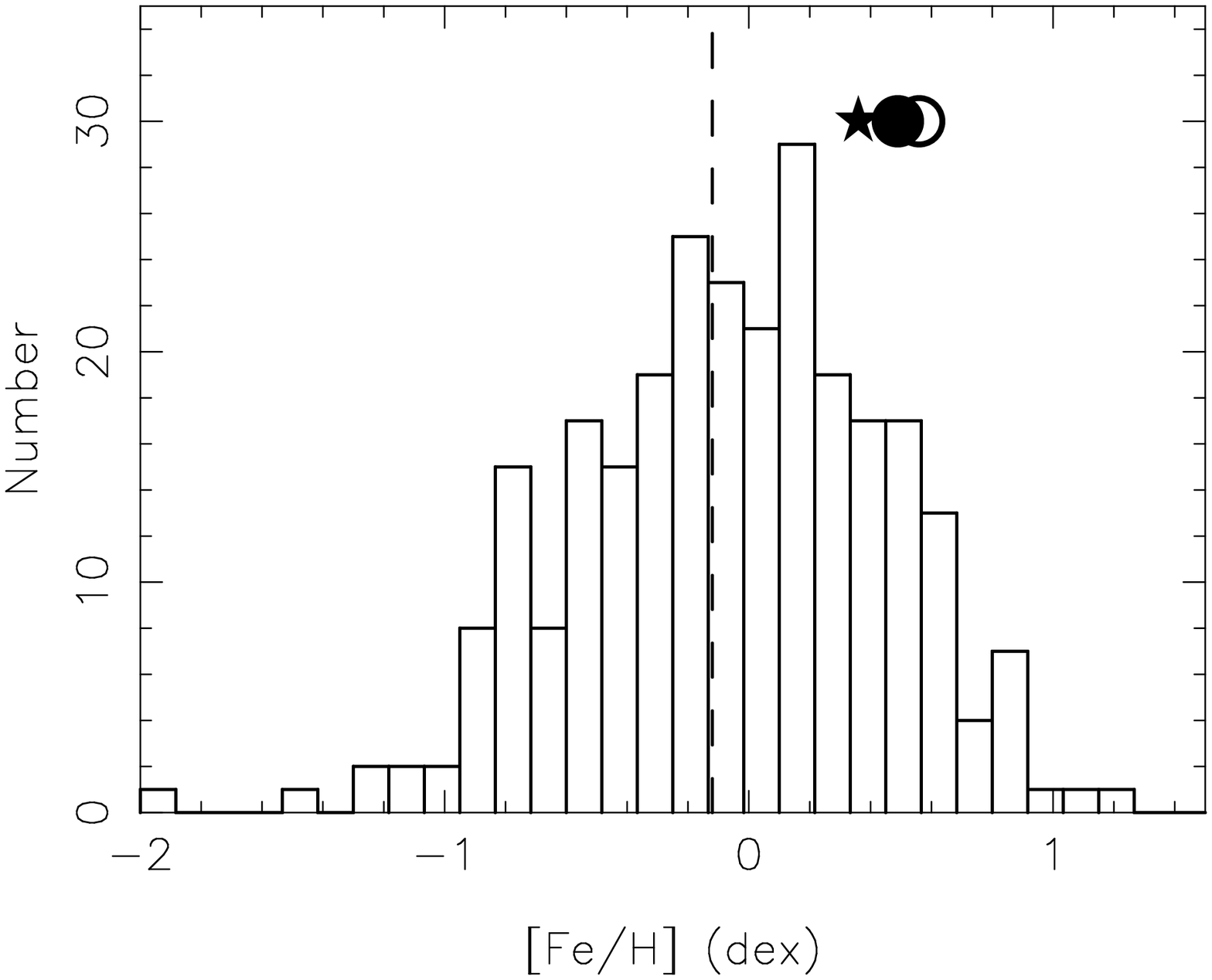}
\caption[]{Fe-metallicity distribution for the sample of 268 Galactic
bulge stars from \cite{sadler96} as re-calibrated by \cite{fulbright06}
is shown.  The median at [Fe/H] $-0.12$~dex is indicated by the dashed vertical line.
The [Fe/H] values for the three 
microlensed bulge dwarfs, \mystar\ studied here (large filled circle),
\johnstar\ \citep{johnson07} (large open circle), and 
\moa\ \citep{johnson08} (star symbol), are marked.
\label{figure_feh_hist}}
\end{figure}

All of these samples of Galactic bulge
stars are of luminous giants and/or of red clump stars.  They all
have mean and median [Fe/H] values that are slightly sub-solar and
are similar to the mean [Fe/H] of $-0.1$~dex found for local disk stars by
\cite{localdisk}.  Yet \mystar, analyzed here, 
and \johnstar\ \citep{johnson07}, another 
Galactic bulge microlensed dwarf with a high
quality detailed abundance analysis, both have
[Fe/H] $\sim +0.5$~dex; the third such star found to date, \moa\, 
with a lower signal-to-noise ratio spectrum, analyzed by
\cite{johnson08}, has a somewhat lower Fe-metallicity,
[Fe/H] = $+0.36\pm0.18$~dex.
The comparison between the [Fe/H] distribution for
the recalibrated sample of \cite{sadler96} by \cite{fulbright06}
and the [Fe/H] values deduced for these three Galactic bulge
dwarfs is shown in Figure~\ref{figure_feh_hist}.  While three stars is 
an uncomfortably small sample,
it is difficult to believe that this
is consistent with the published metallicity distributions for
the Galactic bulge giants and/or red clump stars.
The probability that the sample of three stars would have such
high [Fe/H] values  by chance is less than
1\%  given the  [Fe/H] distributions
of the larger of the relevant  studies 
in the Galactic bulge claiming to have unbiased samples, 
including \cite{ramirez00} with 110 M giants, which must be considered
unbiased at the high metallicity end,
\cite{sadler96} (268 bulge stars),  and the recalibrated verion of the latter
by \cite{fulbright06}.  A KS test  shows that the probability
of finding the three carefully studied microlensed bulge dwarfs at their 
very high  [Fe/H] values if the underlying metallicity distribution
is that found by \cite{zoccali08} from high dispersion spectra
of a very large sample of giants in three bulge fields is
less than 10$^{-4}$.  This calculation takes into account
the radial gradient in the metallicity distribution function
found by \cite{zoccali08}.

%

We next assemble the evidence that \mystar\ is actually
a dwarf in the Galactic bulge.
We have already demonstrated that the microlensed bulge dwarf studied
here has
stellar parameters that, when combined with the unlensed magnitude
from the mirolensing light curve, yield an unlensed luminosity
consistent with that expected for a dwarf of
its \teff\ at the distance
of the Galactic center.  This  is true  to within 0.2~mag for 
the other two such stars as well \johnstar\ and \moa\
\citep{johnson07,johnson08}.
The three microlensed bulge dwarfs studied to date 
come from a kinematically hot population; they have radial velocities
of +99, $-$154 and +113 \kms\ with a typical uncertainty of less
than 2~\kms, thus showing a dispersion in $v_r$ consistent
with that inferred from large samples of bulge giants
\citep[see, e.g.][]{sadler96,zoccali08}.  

About 4\% of the 
dwarfs in the solar neighborhood from the proper motion
sample of \cite{grenon89} appear to be old and very metal-rich,
with [M/H] $> +0.30$~dex.  These stars are almost all on
eccentric orbits with small pericenters ($\lesssim 3$~kpc). 
\cite{castro97} and \cite{pompeia03} have carried out detailed
abundance analyses of some of these stars to find that the most metal rich
of them reach [Fe/H] +0.55~dex, and have solar abundance ratios
in general. These local dwarfs are thus similar in their
abundances to \mystar.  The references cited suggest that
they are possibly stars on chaotic orbits ejected from the
Galactic bar or older central regions of the Galactic disk.
Little is known about the inner regions of the Galactic disk,
whether it exists at all in the inner kpc, what its scale height
might be should it exist, etc. The model of \cite{dirbe_phot}, constructed
to reproduce the surface brightness within 5~kpc
of the Galactic center seen by COBE/DIRBE at 3.5 and at 4.9~$\mu$
after correction for extinction,
suggests a double exponential disk with a scale heights of 42
and of 210~pc (1.5$^{\circ}$ at a distance of 8~kpc to
the Galactic center) combined with a truncated power-law bulge.  However,
the beam size of COBE/DIRBE is $0.7^{\circ} \times 0.7^{\circ}$,
and this was smoothed to a 1.5$^{\circ}$ angular resolution
for their analysis. The
scale height of 210~pc thus corresponds to their minimum
angular resolution, and may well be smaller, which would reduce
any potential disk contamination of bulge samples.  In any case, 
using their model as a guide,
we find that the probability of contamination by disk stars 
of a sample at $b = 2.5^{\circ}$
in this maximal case is only $\sim$30\% larger 
than it is at $b = 4^{\circ}$.  Some disk models such as
model 2 of \cite{2mass_bulge}, constructed to fit 2MASS star counts,
contain an inner hole, which would lower potential disk 
contamination of bulge samples substantially.

Rather than speculate further on whether or not there is a 
disk within the central kpc and what its properties might be, 
we refer back to the large samples
of ``bulge'' giants studied in the many references cited above.
The innermost field included in most of these is Baade's Window,
with $b$ ranging from $-3.9$ to $-4.1^{\circ}$.  
The galactic latitudes for the three microlensed dwarfs are
$-2.5$, $-3.6$ and $-4.8^{\circ}$, so that two of these
are slightly
smaller than that of Baade's window.  However, recently 
\cite{rich07} presented a detailed abundance analysis
from near-IR spectra of M giants in a field at
($l,b$) = ($0^{\circ},-1^{\circ}$), significantly closer
to the Galactic center than any of the three 
microlensed dwarfs.  Their sample of 17 M giants
has the usual properties seen in the Baade's Window
samples, a mean [Fe/H] of $-0.22$~dex, with the most
metal rich at [Fe/H] $+0.02\pm0.11$~dex.  These  giants
show the usual $\alpha$-enhancement.  Surely the
microlensed dwarfs, each located more than twice as high
above the Galactic plane than the giants in this inner
bulge field, then cannot be from a different
 population than the \cite{rich07} M giant sample;  we thus
do not believe that the sample of microlensed bulge stars is
contaminated by any disk stars that might be
co-located near the Galactic center.

We suggest that sample bias is responsible for the difference
between the median Fe-metallicity seen in the bulge giant samples and
that of the microlensed bulge dwarfs.  
A possible mechanism is the very high mass loss rates
predicted to occur  at the very high metallicities and moderately
high luminosities
being discussed here.  These are  high enough that in an old stellar 
population,
stars are predicted to lose enough mass to peel off the 
red giant branch (RGB) before reaching the He-flash,
which they never go through.  Red clump stars, which are burning He
in their cores on the horizontal branch, are even more evolved than giants
at the RGB tip, and so their evolutionary tracks, once
mass loss is properly taken into account, may indicate
that their expected numbers in an old very-metal rich population
may be even more depleted  than are luminous first ascent RGB stars.
This effect may have been detected through CMD studies of 
the extremely metal-rich open cluster NGC~6791 with [Fe/H] +0.45~dex \citep{carretta07}. 
In this cluster, \cite{kalirai07} found a strong relative absence
of luminous RGB stars.
We propose that this relative paucity
of giants on the upper RGB for very metal-rich old populations produces a bias against
the highest metallicity stars by preferentially eliminating them from the
samples being studied by all previous investigations in the Galactic bulge,
which samples consist of one or more of the luminous K giants, 
luminous M giants, and red clump stars found there.

If this is correct, then the mean bulge metallicity may be comparable to that
expected from the radial gradients prevailing within the Galactic disk
at a time $\sim$5~Gyr ago extrapolated to the Galactic center.  It thus may be
that at the present time there
is a gradient
within the solar circle of stellar metallicity with Galactic radius
that is roughly comparable to that measured for the interstellar
medium (ISM) at present
from HII regions by \cite{esteban05} and from planetary nebulae (PN) by
\cite{pn_grad}.  The latter present
an estimate of the change of the radial
abundance gradient in the Galaxy as a function of time from PN of
varying ages.  They
suggest that the radial abundance gradient in the ISM in the Galactic disk
was twice as large $\sim$5~Gyr ago when the Sun was formed
than it is now.  

There are important consequences for the chemical evolution
of extragalactic objects as well
if our conjecture regarding
the metallicity distribution of the Galactic bulge 
is correct.  The interpretation of spectra
and broad band photometry of the integrated light of simple
old metal-rich stellar populations such as are believed to exist
in luminous elliptical galaxies may be affected.  We expect
an underestimate of the true Fe-metallicity of such systems
to occur with commonly used tools such as Lick indices, calculated
with stellar evolutionary tracks and isochrones that ignore
mass loss, as is normally the case.  We expect an overestimate
of the mass-to-light ratio to occur if the luminous RGB
stars we expect to be present  near the RGB tip in a metal-rich population 
are not present
in their expected numbers
as they contribute only a small fraction of the
mass, but a larger fraction of the total luminosity.  
Changes in  the strength of specific absorption features
can be expected as well, particularly in the IR, where luminous
red giants dominate
the integrated light assuming they are present.

\section{Abundance Ratios \label{section_abund_ratios} }

Many previous analyses of Galactic bulge K and M giants
have found enhanced $\alpha$/Fe ratios when
compared to both the thick and thin disk stars for which
very accurate abundance ratios for many elements for large samples
exist, see, e.g., \cite{reddy03} or \cite{bensby05}. 
Abundance ratios help illuminate the star formation
history of the bulge, and the contributions to its chemical
inventory as a function of time of SNIa, SNII, and AGB stars.
The $\alpha$-enhancement found in Galactic bulge giants has been
viewed as an indication that chemical evolution
proceeded more rapidly in the
Galactic bulge than in either the Galactic thick or thin disk population.
In particular, \cite{bulge_chemevol} discuss the role of O and Mg as probes
of the contribution of massive stars; see \cite{ballero}  for a current
model.  \cite{mcwilliam08} revisit this issue, suggesting that metallicity
dependent mass loss rates or nucleosynthesis yields may
be important in this context.
Such findings help to determine the mode of formation of the
Galactic bulge, whether it is a classical bulge or one formed over
a longer timescale through secular evolution of the disk.
Detailed discussion of these options and  references
to the relevant theoretical studies can be found in e.g.,
\cite{lecureur07}.

Rapidly increasing samples of luminous stars in the bulge, both in terms of 
numbers of stars
analyzed and in terms of the accuracy of the results, are now
becoming available.
\cite{fulbright07} gives a detailed discussion of $\alpha$-enhancement
in their sample of Galactic bulge K giants.  They
find enhanced [O/Fe], [Mg/Fe] and [Al/Fe] 
compared both
to the solar ratios and to those in both thick and thin disk stars 
even in super-solar bulge K giants and discuss the different behavior
of the hydrostatic (including O and Mg) and explosive 
(including Si, Ca and Ti) $\alpha$ elements; only the
former appear enhanced in the bulge giants.
Recently initial results from a major high dispersion spectroscopic
survey of Galactic bulge giants and red clump stars being carried out
at the VLT have been reported by \cite{zoccali06} with more details in
\cite{lecureur07}.  They  present
the   analysis of a sample of 53 stars selected to span the
color range of bulge giants.

\begin{figure}
\epsscale{1.0}
\plotone{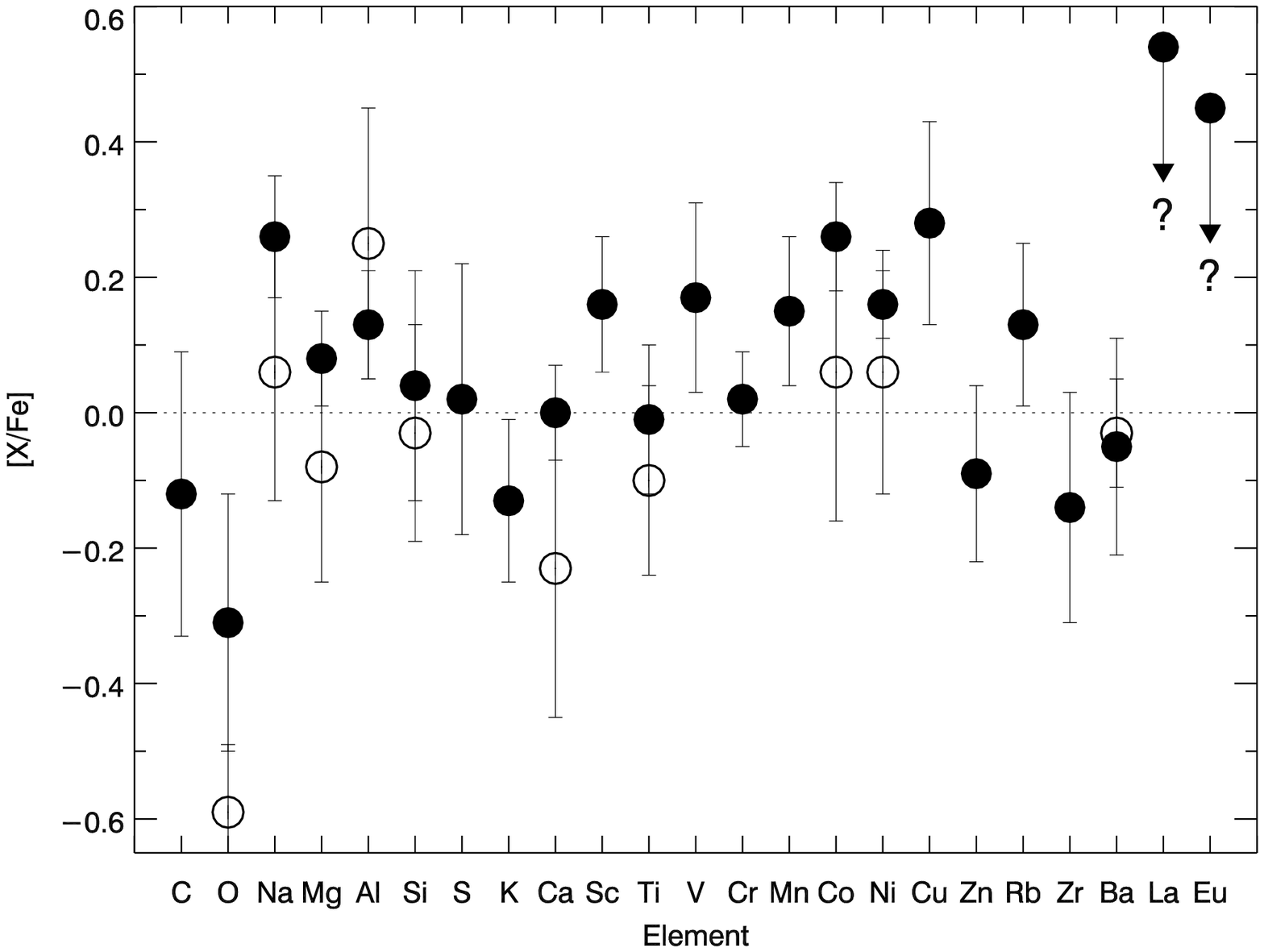}
\caption[]{Abundance ratios  for various species are shown
for \mystar\ analyzed here (filled circles) and for \johnstar\ (open circles)
from \cite{johnson07}.
\label{figure_two_ogle}}
\end{figure}

\begin{figure}
\epsscale{0.85}
\plotone{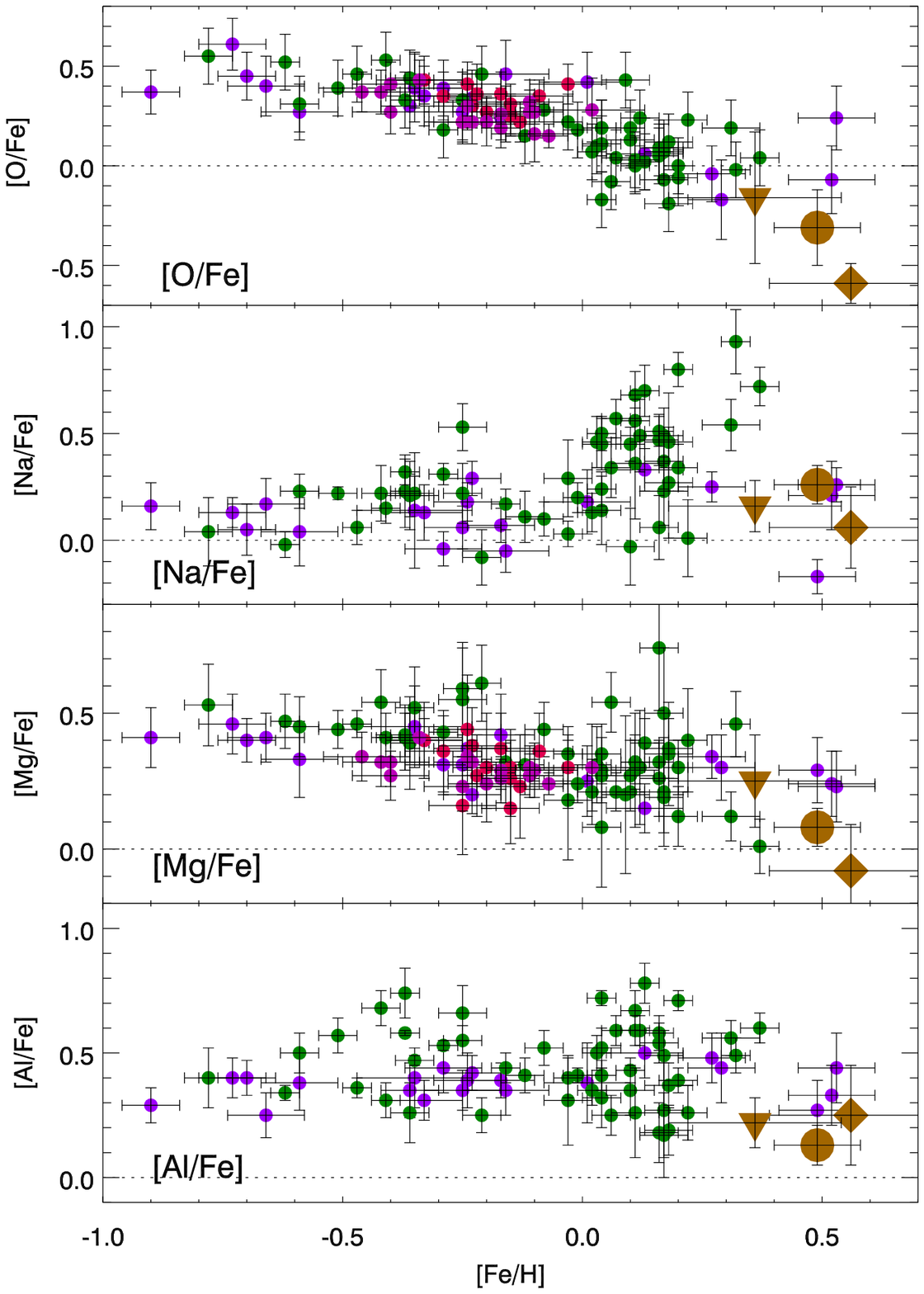}
\caption[]{Abundance ratios
[O/Fe], [Na/Fe], [Mg/Fe] and [Al/Fe] are shown as a function
of [Fe/H].  The three microlensed stars are indicated by
lage brown symbols: \mystar\ (filled circle),
\johnstar\ from \cite{johnson07} (diamond),
and  \moa\ from \cite{johnson08} (inverted triangle).  They are shown
superposed on those for samples
of bulge M and K giants of \cite{fulbright07} (blue circles), \cite{rich05}
(red circles), \cite{lecureur07} (green circles),
and for M giants in the inner bulge from \cite{rich07} 
(pink circles).
\label{figure_abund_ratios1}}
\end{figure}

\begin{figure}
\epsscale{1.0}
\plotone{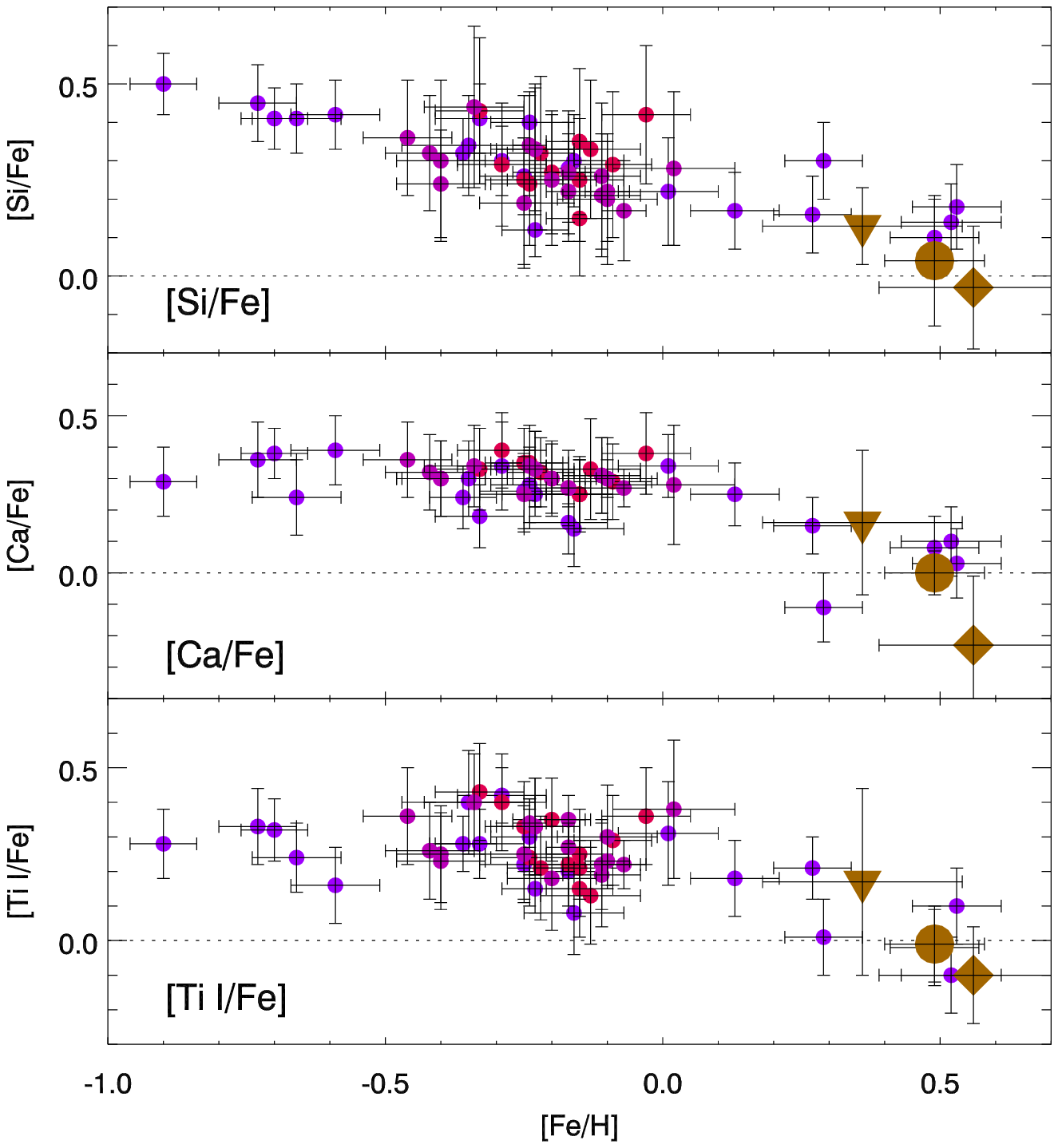}
\caption[]{The same as Fig.~\ref{figure_abund_ratios1} for
[Si/Fe], [Ca/Fe] and [Ti/Fe].  The symbols are the same
as in Fig.~\ref{figure_abund_ratios1}.
\label{figure_abund_ratios2}}
\end{figure}

However, our analysis of \mystar\ (see Table~\ref{table_abunds}) finds 
[Mg/Fe] has the solar
ratio and [O/Fe] is subsolar.  \cite{johnson07} find similar low ratios in
\johnstar.  The third microlensed star, \moa, shows a 
small $\alpha$-enhancement,
but the uncertainties are large, and a  $1\sigma$ deviation toward
smaller values would make them
consistent with the solar values.  
Subsolar [O/Fe] is found in all three microlensed dwarfs. 
Figure~\ref{figure_two_ogle}
shows the abundance ratios [X/Fe] for each species detected
in one of both of the two bulge dwarfs with [Fe/H] $\sim +0.5$~dex.
Figures~\ref{figure_abund_ratios1}
and \ref{figure_abund_ratios2} display the abundance ratios
for the three microlensed Galactic bulge dwarfs
for several elements as a function of
[Fe/H] as compared to recent results from samples
of luminous Galactic bulge giants.  

\begin{figure}
\epsscale{0.9}
\plotone{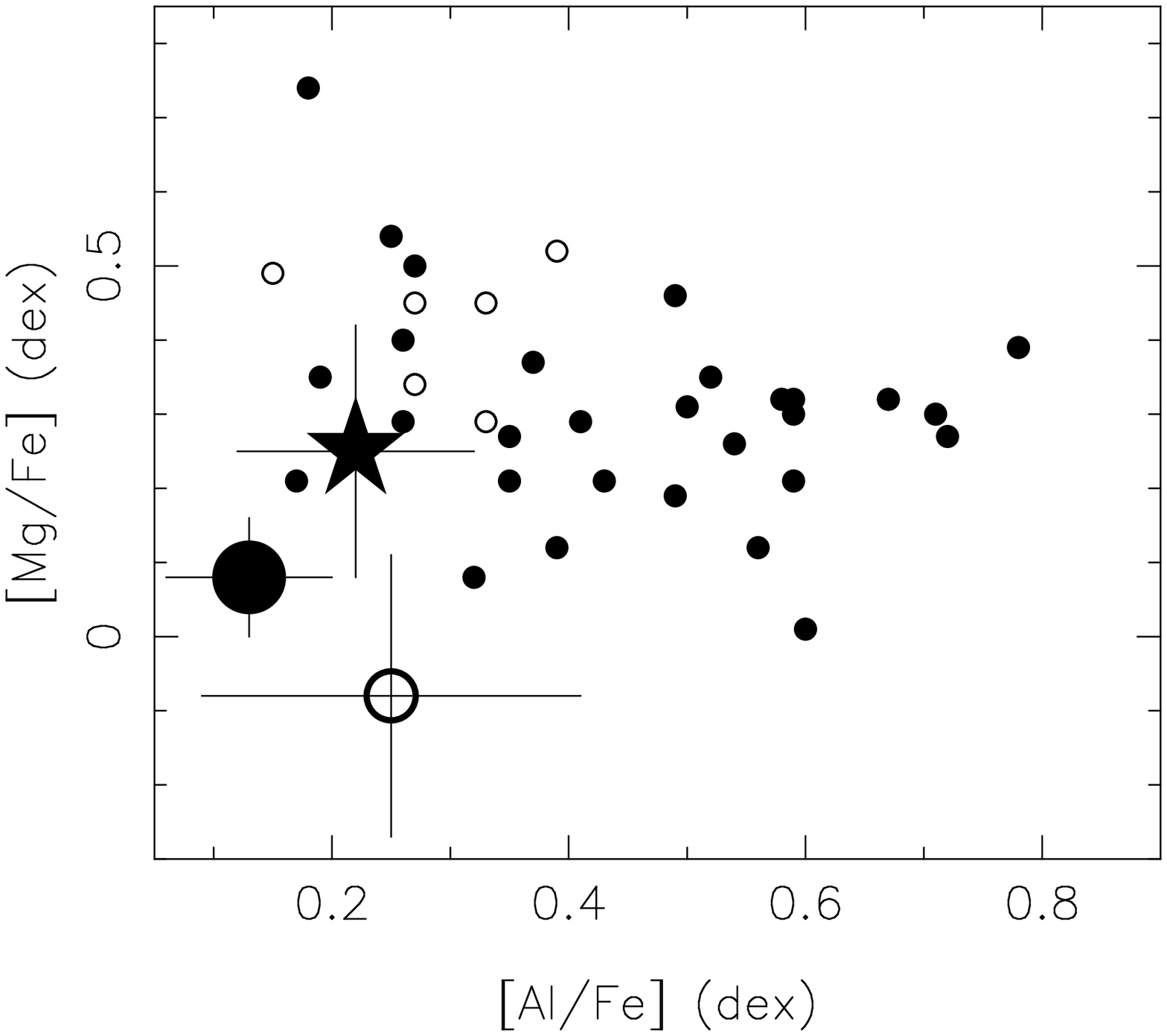}
\caption[]{The [Mg/Fe] ratio versus the [Al/Fe] ratio for \mystar,
analyzed here, 
is shown as the large filled circle,
 \johnstar\ from \cite{johnson07}  as the large open circle,
and \moa\ from \cite{johnson08} as the large star. 
Those stars with [Fe/H] $> 0.0$~dex from the samples of bulge
giants and red clump stars of  \cite{lecureur07}
are shown as small filled circles,
and those from the sample of \cite{fulbright07} 
 as small open circles.   Error
bars (1$\sigma$) are shown for the three OGLE dwarfs; those of 
the two samples of bulge giants
are similar to those for \johnstar\ taken from \cite{johnson07}.
\label{figure_mgal}}
\end{figure}

A comparison of [Mg/Fe] with [Al/Fe] for the two microlensed
bulge dwarfs compared to the results from the samples of
bulge giants of \cite{lecureur07} and of \cite{fulbright07} 
is shown in Figure~\ref{figure_mgal}. 
The three microlensed bulge dwarfs lie at the low end of the
distribution in [Mg/Fe] and at the extreme low end of that
for [Al/Fe]. Also intriguing is
that \cite{lecureur07} finds evidence for 
variations among the giants of highest
metallicity in their sample of [O/Fe], [Na/Fe], [Mg/Fe] and [Al/Fe]
that substantially exceed their claimed uncertainties.  This is
accompanied by an anti-correlation between [O/Fe] and
[Na/Fe] and by a correlation of [Na/Fe] with [Al/Fe], very reminiscent
of the star-to-star abundance variations found within
Galactic globular clusters, as summarized in the 
recent review of \cite{gratton_araa}.


As is shown in Figs.~\ref{figure_abund_ratios1} and \ref{figure_abund_ratios2},
the microlensed dwarfs show no sign of $\alpha$-enhancement,
while the bulge giants have [Mg/Fe] (as well as [Na/Fe] and [Al/Fe])
enhanced even at super-solar metallicities.
What is causing this difference ?    Why do the giants show
a such large range in [X/Fe] for several key light elements ?
Are the giants mixed ?  Are there substantial non-LTE corrections which
are not being included correctly  ?

The key question 
is which of the two, bulge giants or microlensed bulge
dwarfs, best represents in its atmosphere today the
initial chemical
inventory of the star, i.e., of the interstellar medium at 
the time that the star
was formed.  It must be noted that the study of microlensed Galactic
dwarfs can be carried out differentially with respect to the Sun, as we did here,
while those of cool giants are forced to 
rely on Arcturus and/or $\mu$~Leo as the reference. This 
difference may give rise to
potentially  substantial systematic errors in the analysis
 which may depend
on \teff\ of the star.  A probable example of such a systematic
error can be seen in the [Si/Fe] panel of Figure~3 of \cite{preston},
where this abundance ratio, inferred in every star from the
same line of Si~I, at 3905~\AA,
changes systematically from
about +0.7 to 0.0~dex as \teff\ increases from 5000 to 6500~K
for a sample of very metal-poor RHB stars and red giants.
Presumably [Si/Fe] is approximately the same for all the
stars included in this figure, but some 
unrecognized systematic error has crept
into the analyses so as to produce the strong trend they found for [Si/Fe]
with \teff.

It must be recognized, however, that the recent analyses of bulge K and M giants
have been carried out by several independent groups each of which contains
highly experienced experts in this field.  We therefore look for an explanation
that allows both sets of deduced abundance ratios, those of the bulge giants
and those found here for \mystar\ as well as for the other two
well studied microlensed bulge dwarfs, to be correct.
Dredge up and mixing in the giants is a possible culprit,
particularly since \cite{lecureur07} found a large range in
[O/Fe], [Na/Fe], [Mg/Fe] and [Al/Fe] among the most Fe-rich
stars in their large sample of bulge giants.
However, at least
in the globular cluster case 
at low [Fe/H], one sees C burning into N, and sometimes
O burning into N as well \citep{m5_cn}, occasionally
with depleted Mg and enhanced Al.  The signature of proton burning
via the CN cycle, clearly seen in the globular cluster giants,
is not apparent in the bulge giants.  For this (and other)
reasons, \cite{lecureur07} rejected
mixing as the explanation for the large range they saw in 
in some abundance ratios. 

If  larger samples of microlensed Galactic bulge dwarfs
continue to show solar  ratios of
[Mg/Fe] and [Al/Fe] while samples of bulge giants show
in the mean much higher values for each of these elements
coupled with a large range for each, even at super-solar metallicity,
we would suggest that the true  initial
abundance ratios for the light elements,
including the $\alpha$-elements, 
at high [Fe/H] in the Galactic bulge are those of the dwarfs, not the
giants.

\section{Summary \label{section_summary}  }

We have analyzed a  high dispersion 
spectrum of a microlensed dwarf, \mystar, in the Galactic bulge.  The
magnification of this event was very high, HIRES on the 10-m
Keck~I Telescope was used, the weather was clear with good seeing,
and the exposure
time was long compared to any previous such data, so
the resulting spectrum has a relatively  high signal-to-noise ratio.

We stress that in principle the abundance analysis of
a upper main sequence dwarf is much easier and less prone to error
for spectra of a fixed signal-to-noise ratio
than that of a much cooler but much brighter bulge giant with a very
complex spectrum full of blends and of strong molecular bands.
The advantages of analyzing microlensed bulge stars, for which
the required high signal-to-noise ratio can sometimes be achieved, to 
improving our understanding of
the [Fe/H] distribution and chemical evolution of the Galactic bulge
are large.  

We have derived  for \mystar, which we believe to be
a dwarf below the main sequence turnoff with \teff $\sim 5400$~K,
a very high Fe-metallicity, [Fe/H] = \fehogle ~dex.  This is very peculiar given that many previous surveys
of the metallicity distribution carried out with
large samples of K or M giants in the Galactic bulge find both the
mean and median [Fe/H] to be sub-solar.  The two other 
highly magnified Galactic bulge dwarfs studied in detail, \johnstar\
by \cite{johnson07} and \moa\ by \cite{johnson08}, also have very 
high  Fe-metallicities.

In order to produce consistency, we suggest that there is a sampling
bias in the bulge giant samples such that very metal-rich giants
are strongly depleted.  We suggest a physical mechanism 
for this, the very high mass loss rates expected for such
metal-rich old giants can exhaust their envelopes prior to the
normal He-flash.

We also find that \mystar\ does not show enhancements of
the $\alpha$ elements; neither does \johnstar, analyzed by
\cite{johnson07}.  However, most bulge giants
from samples with high dispersion spectroscopy, e.g. 
the work of \cite{fulbright07} and particularly those from
\cite{lecureur07}, do show large (and varying from star to star)  
enhancements of [Na/Fe],  [Mg/Fe]
and [Al/Fe] even at super-solar metallicities.
We suggest that it is the abundances deduced for the microlensed
dwarfs that best represent the initial chemical inventory
of the interstellar medium at the time these stars formed, while
those derived for the bulge giants may not.

We recognize that three stars is a very small sample, but the
implications of our results and inferences for the chemical
evolution of the Galactic bulge and for the interpretation
of integrated light spectra and broad-band photometry
of old simple stellar populations such as  luminous
ellitical galaxies, are so important that we offer
these hypotheses at this time.

The study of additional highly microlensed Galactic bulge dwarfs
 to increase the sample from just three such stars is clearly urgent.
Now that ongoing microlensing surveys make such observations
of Galactic bulge dwarfs feasible,   we expect substantial
improvements in the sample size of high quality spectra
for Galactic bulge dwarfs. 
Suitable
high magnification events are  rare and lining up the necessary
instruments/telescopes/clear weather at just the right time is difficult.
Several years may be required to accumulate a suitable sample
of spectra of highly microlensed dwarf stars in the Galactic bulge.
We are  now appropriately
positioned to carry out such a time critical program at the Keck Observatory
over the next few years, and eagerly look forward to
confirmation of our perhaps premature and provocative
hypotheses in the not too distant future.

\acknowledgements

We are grateful to the many people  
who have worked to make the Keck Telescope and HIRES  
a reality and to operate and maintain the Keck Observatory. 
The authors wish to extend special thanks to those of Hawaiian ancestry
on whose sacred mountain we are privileged to be guests. 
Without their generous hospitality, none of the observations presented
herein would have been possible.

J.G.C. and W.H. are grateful to NSF grant AST-0507219 to JGC for partial
support.  Work by A.G. was supported by NSF grant AST-042758.
The OGLE project is partially supported by the Polish MNiSW grant
N20303032/4275.  We thank Manuela Zoccali for providing 
results in advance of publication
on the metallicity distribution function of bulge giants.

\section{Appendix A -- Diffuse Interstellar Bands \label{appendix_dib} }

The origin of the diffuse interstellar bands (DIBs) has been a puzzle
for more than 30 years; see, e.g., the review by \cite{herbig}.
Most stars previously studied with high reddening are  hot luminous
young stars within a cluster embedded
in a single cloud complex whose column density
varies somewhat over the angular extent of the cloud, such as the
IV Cyg association.
But the reddening seen for \mystar\ ($E(B-V) = 0.68\pm0.10$~mag), while not 
particularly large,
results from the
many clouds along the line of  sight to
the Galactic center, and hence is much more representative
of the typical ISM.
Very little, if any, of this reddening is believed to arise
arise within the Galactic bulge itself.  

Because \mystar\ is
a metal-rich cool dwarf, rather than an O or B star which would is normally
used for such work, measurements of the DIBs
are difficult and the results more uncertain than would
arise for a  much less complex hot star spectrum.   The very strong and blended
stellar absorption features
in the blue part of the spectrum made a clear detection
of the classical DIB at 4430~\AA\ impossible. Parameters
for those interstellar bands that could be detected with certainty
are given in Table~\ref{table_dib}.  They are comparable in strength
with those observed for the well studied B7 supergiant HD~183143
whose values  are given in Table~A1 of \cite{herbig} even though
the reddening of this bright star is $E(B-V) = 1.28$~mag, about
twice as large as that of \mystar.  \cite{snow76} have shown
that the band strength per grain in the line of sight
apparently decreases with increasing grain size, such that 
dense interstellar clouds are less efficient in producing
absorption by DIBs for the same total reddening than are
less dense, but more numerous clouds in the line of sight.  The
very strong DIBs and immense interstellar absorption in the NaD lines
in our spectrum of \mystar\  should further elucidate this ``skin'' effect.


{}

\clearpage

%
%
\begin{deluxetable}{lcrrr }
\tablenum{1}
\tablewidth{0pt}
\tablecaption{$W_{\lambda}$ for \mystar \label{table_eqw}}
\tablehead{
\colhead{$\lambda$} & \colhead{Species} & \colhead{EP} &
\colhead{log($gf$)} & \colhead{\eqw} \\
\colhead{(\AA)} & \colhead{} &
\colhead{(eV)}    &   \colhead{(dex)}  & \colhead{(m\AA)} 
}
\startdata 
 6707.76 & Li~I &   0.00 &   0.178 &     $\leq$5.0  \\
 7111.45 & C~I &    8.64  & $  -1.000 $ &    21.2  \\
 7115.32 & C~I &    8.64  & $  -0.600 $ &    27.0  \\
 7116.96 & C~I &    8.64  & $  -1.000 $ &    21.8  \\
 6300.30 & [O~I] &    0.00  & $  -9.720 $ &  $\leq$8.6  \\
 7771.94 & O~I &   9.15 &   0.369 &    52.2  \\
 7774.17 & O~I &   9.15 &   0.223 &    54.6  \\
 7775.39 & O~I &   9.15 &   0.001 &    40.8  \\
 5682.63 & Na~I  &    2.10  & $  -0.700 $ &   194.1  \\
 5688.19 & Na~I  &    2.10  & $  -0.420 $ &   214.1  \\
 6154.23 & Na~I  &    2.10  & $  -1.530 $ &   105.3  \\
 6160.75 & Na~I  &    2.00  & $  -1.230 $ &   125.8  \\
 5711.09 & Mg~I  &    4.34  & $  -1.670 $ &   168.8  \\
 6318.72 & Mg~I  &    5.11  & $  -2.100 $ &    83.9  \\
 6319.24 & Mg~I  &    5.11  & $  -2.320 $ &    66.2  \\
 6965.41 & Mg~I  &    5.75  & $  -2.000$\tablenotemark{a}
    &    72.3  \\
 6696.02 & Al~I  &    3.14  & $  -1.340 $ &    90.0  \\
 6698.67 & Al~I  &    3.14  & $  -1.640 $ &    66.0  \\
 5701.10 & Si~I  &    4.93  & $  -2.050 $ &    64.3  \\
 6145.02 & Si~I  &    5.61  & $  -1.440 $ &    63.9  \\
 6155.13 & Si~I  &    5.62  & $  -0.760 $ &   126.9  \\
 7235.33 & Si~I  &    5.61  & $  -1.310$\tablenotemark{a}
       &   61.2  \\
 7235.82 & Si~I  &    5.61  & $  -1.590$\tablenotemark{a}
       &    48.0  \\
 7800.00 & Si~I  &    6.18  & $  -0.680$\tablenotemark{a}
      &   101.2  \\
 6756.96 & S~I &    7.87  & $ -0.90 $ &    syn  \\
 6757.15 & S~I &    7.87  & $ -0.31 $ &    syn  \\
 7698.97 & K~I   &    0.00  & $  -0.168 $ &   240.0  \\
 5590.11 & Ca~I  &    2.52  & $  -0.710 $ &   123.0  \\
 5867.56 & Ca~I  &    2.93  & $  -1.340$\tablenotemark{a}
       &    60.0  \\
 6156.02 & Ca~I  &    2.52  & $  -2.190 $ &    30.8  \\
 6161.30 & Ca~I  &    2.52  & $  -1.030 $ &   104.5  \\
 6166.44 & Ca~I  &    2.52  & $  -1.050 $ &   107.6  \\
 6169.04 & Ca~I  &    2.52  & $  -0.540 $ &   144.7  \\
 6455.60 & Ca~I  &    2.52  & $  -1.360 $ &    97.1  \\
 6464.68 & Ca~I  &    2.52  & $  -2.150$\tablenotemark{a}
   &  47.7 \\ 
 6471.66 & Ca~I  &    2.52  & $  -0.590 $ &   124.8  \\
 6499.65 & Ca~I  &    2.54  & $  -0.590 $ &   130.3  \\
 6798.48 & Ca~I  &    2.71  & $  -2.420$\tablenotemark{a}
      &    31.5  \\
 5684.20 & Sc~II &    1.51  & $  -1.080 $ &    54.5  \\
 6245.64 & Sc~II &    1.51  & $  -1.130 $ &    49.5  \\
 5453.64 & Ti~I  &    1.44  & $  -1.610 $ &    27.0  \\
 5648.57 & Ti~I  &    2.49  & $  -0.252 $ &    44.5  \\
 5739.46 & Ti~I  &    2.25  & $  -0.602 $ &    35.1  \\
 5766.33 & Ti~I  &   3.29 &   0.360 &    35.8  \\
 5913.73 & Ti~I  &    0.02  & $  -3.780$\tablenotemark{a}
        &     8.0  \\
 5918.54 & Ti~I  &    1.07  & $  -1.470 $ &    52.7  \\
 6092.80 & Ti~I  &    1.89  & $  -1.380 $ &    21.5  \\
 6716.67 & Ti~I  &    2.49  & $  -1.060$\tablenotemark{a}
        &     9.7  \\
 6606.95 & Ti~II &    2.06  & $  -2.790 $ &    18.6  \\
 5670.85 & V~I   &    1.08  & $  -0.425 $ &    68.6  \\
 6081.44 & V~I   &    1.05  & $  -0.579 $ &    67.6  \\
 6090.22 & V~I   &    1.08  & $  -0.062 $ &    90.4  \\
 6251.82 & V~I   &    0.29  & $  -1.340 $ &    81.7  \\
 6285.14 & V~I   &    0.28  & $  -1.510 $ &    52.4  \\
 5702.32 & Cr~I  &    3.45  & $  -0.667 $ &    66.0  \\
 5783.09 & Cr~I  &    3.32  & $  -0.500 $ &    61.0  \\
 5783.89 & Cr~I  &    3.32  & $  -0.295 $ &    84.9  \\
 5787.96 & Cr~I  &    3.32  & $  -0.083 $ &    73.2  \\
 5844.59 & Cr~I  &    3.01  & $  -1.760 $ &    21.0  \\
 6978.49 & Cr~I  &    3.46  &     0.143   &   127.0  \\
 6979.80 & Cr~I  &    3.46  & $  -0.411 $ &    73.4  \\
 5537.74 & Mn~I  &    2.19  & $  -2.020 $ &   133.3  \\
 6013.50 & Mn~I  &    3.07  & $  -0.252 $ &   145.9  \\
 6021.80 & Mn~I  &   3.08 &   0.034 &   163.3  \\
 5916.25 & Fe~I  &    2.45  & $  -2.910 $ &    91.2  \\
 5927.79 & Fe~I  &    4.65  & $  -0.990 $ &    66.2  \\
 5929.67 & Fe~I  &    4.55  & $  -1.310 $ &    70.5  \\
 5930.17 & Fe~I  &    4.65  & $  -0.140 $ &   131.6  \\
 5934.65 & Fe~I  &    3.93  & $  -1.070 $ &   112.2  \\
 5940.99 & Fe~I  &    4.18  & $  -2.050 $ &    42.8  \\
 5952.72 & Fe~I  &    3.98  & $  -1.340 $ &   100.8  \\
 5956.69 & Fe~I  &    0.86  & $  -4.500 $ &    88.8  \\
 5976.79 & Fe~I  &    3.94  & $  -1.330 $ &    98.0  \\
 5983.69 & Fe~I  &    4.55  & $  -0.660 $ &   101.6  \\
 5984.83 & Fe~I  &    4.73  & $  -0.260 $ &   130.4  \\
 6024.05 & Fe~I  &   4.55 &   0.030 &   161.0  \\
 6027.05 & Fe~I  &    4.07  & $  -1.090 $ &    89.2  \\
 6055.99 & Fe~I  &    4.73  & $  -0.370 $ &    99.5  \\
 6065.48 & Fe~I  &    2.61  & $  -1.410 $ &   179.7  \\
 6078.50 & Fe~I  &    4.79  & $  -0.330 $ &   112.2  \\
 6079.00 & Fe~I  &    4.65  & $  -1.020 $ &    73.1  \\
 6089.57 & Fe~I  &    5.02  & $  -0.900 $ &    61.8  \\
 6093.67 & Fe~I  &    4.61  & $  -1.400 $ &    54.7  \\
 6094.37 & Fe~I  &    4.65  & $  -1.840 $ &    48.0  \\
 6096.66 & Fe~I  &    3.98  & $  -1.830 $ &    70.1  \\
 6120.25 & Fe~I  &    0.92  & $  -5.970 $ &    24.3  \\
 6127.90 & Fe~I  &    4.14  & $  -1.400 $ &    76.1  \\
 6136.99 & Fe~I  &    2.20  & $  -2.950 $ &   109.5  \\
 6151.62 & Fe~I  &    2.18  & $  -3.370 $ &    80.3  \\
 6157.73 & Fe~I  &    4.07  & $  -1.160 $ &    92.4  \\
 6159.37 & Fe~I  &    4.61  & $  -1.920 $ &    34.9  \\
 6165.36 & Fe~I  &    4.14  & $  -1.470 $ &    71.8  \\
 6173.34 & Fe~I  &    2.22  & $  -2.880 $ &   107.1  \\
 6180.20 & Fe~I  &    2.73  & $  -2.650 $ &    92.4  \\
 6187.99 & Fe~I  &    3.94  & $  -1.620 $ &    87.2  \\
 6200.31 & Fe~I  &    2.61  & $  -2.370 $ &    97.5  \\
 6232.64 & Fe~I  &    3.65  & $  -1.220 $ &   128.9  \\
 6240.65 & Fe~I  &    2.22  & $  -3.170 $ &    81.9  \\
 6246.32 & Fe~I  &    3.60  & $  -0.880 $ &   177.9  \\
 6252.55 & Fe~I  &    2.40  & $  -1.770 $ &   174.2  \\
 6265.13 & Fe~I  &    2.18  & $  -2.540 $ &   130.6  \\
 6271.28 & Fe~I  &    3.33  & $  -2.700 $ &    52.0  \\
 6290.97 & Fe~I  &    4.73  & $  -0.730 $ &   108.6  \\
 6297.79 & Fe~I  &    2.22  & $  -2.640 $ &   116.0  \\
 6301.51 & Fe~I  &    3.65  & $  -0.718 $ &   181.6  \\
 6302.50 & Fe~I  &    3.69  & $  -1.110 $ &   126.0  \\
 6311.50 & Fe~I  &    2.83  & $  -3.140 $ &    71.6  \\
 6315.31 & Fe~I  &    4.14  & $  -1.230 $ &   111.4  \\
 6315.81 & Fe~I  &    4.07  & $  -1.610 $ &    71.4  \\
 6322.69 & Fe~I  &    2.59  & $  -2.430 $ &   121.3  \\
 6336.82 & Fe~I  &    3.69  & $  -0.856 $ &   167.1  \\
 6355.03 & Fe~I  &    2.84  & $  -2.290 $ &   108.5  \\
 6380.75 & Fe~I  &    4.19  & $  -1.380 $ &    85.7  \\
 6392.54 & Fe~I  &    2.28  & $  -3.990 $ &    49.6  \\
 6408.03 & Fe~I  &    3.69  & $  -1.020 $ &   155.0  \\
 6430.84 & Fe~I  &    2.18  & $  -1.950 $ &   175.5  \\
 6436.41 & Fe~I  &    4.19  & $  -2.450 $ &    34.6  \\
 6475.63 & Fe~I  &    2.56  & $  -2.940 $ &    97.9  \\
 6481.87 & Fe~I  &    2.28  & $  -3.010 $ &   105.7  \\
 6483.94 & Fe~I  &    1.48  & $  -5.340 $ &    18.7  \\
 6495.74 & Fe~I  &    4.83  & $  -0.840 $ &    66.5  \\
 6498.94 & Fe~I  &    0.96  & $  -4.690 $ &    90.5  \\
 6518.37 & Fe~I  &    2.83  & $  -2.450 $ &    88.9  \\
 6533.93 & Fe~I  &    4.56  & $  -1.360 $ &    71.3  \\
 6546.24 & Fe~I  &    2.76  & $  -1.540 $ &   154.3  \\
 6581.21 & Fe~I  &    1.48  & $  -4.680 $ &    63.2  \\
 6592.91 & Fe~I  &    2.73  & $  -1.470 $ &   165.9  \\
 6593.87 & Fe~I  &    2.43  & $  -2.370 $ &   124.8  \\
 6597.56 & Fe~I  &    4.79  & $  -0.970 $ &    72.6  \\
 6608.02 & Fe~I  &    2.28  & $  -3.930 $ &    49.2  \\
 6609.11 & Fe~I  &    2.56  & $  -2.660 $ &   109.2  \\
 6625.02 & Fe~I  &    1.01  & $  -5.370 $ &    54.8  \\
 6627.54 & Fe~I  &    4.55  & $  -1.580 $ &    61.8  \\
 6633.75 & Fe~I  &    4.79  & $  -0.800 $ &   102.0  \\
 6646.93 & Fe~I  &    2.61  & $  -3.960 $ &    38.5  \\
 6653.91 & Fe~I  &    4.15  & $  -2.520 $ &    24.3  \\
 6699.15 & Fe~I  &    4.59  & $  -2.100 $ &    24.1  \\
 6703.57 & Fe~I  &    2.76  & $  -3.060 $ &    73.2  \\
 6710.32 & Fe~I  &    1.48  & $  -4.870 $ &    53.2  \\
 6713.77 & Fe~I  &    4.79  & $  -1.500 $ &    45.0  \\
 6715.38 & Fe~I  &    4.61  & $  -1.540 $ &    73.1  \\
 6716.22 & Fe~I  &    4.58  & $  -1.850 $ &    51.5  \\
 6725.35 & Fe~I  &    4.19  & $  -2.250 $ &    39.9  \\
 6726.67 & Fe~I  &    4.61  & $  -1.070 $ &    78.0  \\
 6733.15 & Fe~I  &    4.64  & $  -1.480 $ &    53.4  \\
 6739.52 & Fe~I  &    1.56  & $  -4.790 $ &    42.2  \\
 6745.11 & Fe~I  &    4.58  & $  -2.170 $ &    27.8  \\
 6746.95 & Fe~I  &    2.61  & $  -4.300 $ &    15.4  \\
 6750.15 & Fe~I  &    2.42  & $  -2.580 $ &   116.5  \\
 6752.71 & Fe~I  &    4.64  & $  -1.200 $ &    78.2  \\
 6806.86 & Fe~I  &    2.73  & $  -3.210 $ &    75.1  \\
 6837.02 & Fe~I  &    4.59  & $  -1.690 $ &    39.4  \\
 6839.83 & Fe~I  &    2.56  & $  -3.350 $ &    67.2  \\
 6842.68 & Fe~I  &    4.64  & $  -1.220 $ &    68.8  \\
 6843.65 & Fe~I  &    4.55  & $  -0.830 $ &    94.3  \\
 6851.63 & Fe~I  &    1.61  & $  -5.280 $ &    21.3  \\
 6855.18 & Fe~I  &    4.56  & $  -0.740 $ &   105.8  \\
 6855.71 & Fe~I  &    4.61  & $  -1.780 $ &    43.5  \\
 6858.15 & Fe~I  &    4.61  & $  -0.930 $ &    77.4  \\
 6861.95 & Fe~I  &    2.42  & $  -3.850 $ &    60.2  \\
 6862.49 & Fe~I  &    4.56  & $  -1.470 $ &    60.1  \\
 6971.93 & Fe~I  &    3.02  & $  -3.340 $ &    41.2  \\
 6978.85 & Fe~I  &    2.48  & $  -2.450 $ &   114.6  \\
 6988.52 & Fe~I  &    2.40  & $  -3.560 $ &    71.3  \\
 6999.88 & Fe~I  &    4.10  & $  -1.460 $ &    94.5  \\
 7000.62 & Fe~I  &    4.14  & $  -2.390 $ &    50.2  \\
 7007.96 & Fe~I  &    4.18  & $  -1.960 $ &    64.8  \\
 7014.98 & Fe~I  &    2.45  & $  -4.200 $ &    31.0  \\
 7022.95 & Fe~I  &    4.19  & $  -1.150 $ &    98.7  \\
 7038.22 & Fe~I  &    4.22  & $  -1.200 $ &   109.0  \\
 7112.17 & Fe~I  &    2.99  & $  -3.000 $ &    79.6  \\
 7114.55 & Fe~I  &    2.69  & $  -4.000 $ &    27.7  \\
 7130.92 & Fe~I  &    4.22  & $  -0.750 $ &   147.5  \\
 7132.98 & Fe~I  &    4.07  & $  -1.630 $ &    71.4  \\
 7142.52 & Fe~I  &    4.95  & $  -1.030 $ &    80.7  \\
 7179.99 & Fe~I  &    1.48  & $  -4.750 $ &    60.7  \\
 7189.15 & Fe~I  &    3.07  & $  -2.770 $ &    82.0  \\
 7285.27 & Fe~I  &    4.61  & $  -1.660 $ &    54.7  \\
 7306.56 & Fe~I  &    4.18  & $  -1.690 $ &    79.9  \\
 7401.69 & Fe~I  &    4.19  & $  -1.350 $ &    70.2  \\
 7418.67 & Fe~I  &    4.14  & $  -1.380 $ &    84.5  \\
 7440.92 & Fe~I  &    4.91  & $  -0.720 $ &    92.9  \\
 7443.02 & Fe~I  &    4.19  & $  -1.780 $ &    73.2  \\
 7447.40 & Fe~I  &    4.95  & $  -1.090 $ &    65.3  \\
 7454.00 & Fe~I  &    4.19  & $  -2.370 $ &    41.2  \\
 7461.52 & Fe~I  &    2.56  & $  -3.530 $ &    66.1  \\
 7563.02 & Fe~I  &    4.83  & $  -1.660 $ &    44.0  \\
 7568.91 & Fe~I  &    4.28  & $  -0.940 $ &   121.7  \\
 7583.79 & Fe~I  &    3.02  & $  -1.890 $ &   128.0  \\
 7586.04 & Fe~I  &    4.31  & $  -0.130 $ &   185.1  \\
 7710.36 & Fe~I  &    4.22  & $  -1.110 $ &   108.0  \\
 7723.30 & Fe~I  &    2.28  & $  -3.610 $ &    86.9  \\
 7748.27 & Fe~I  &    2.95  & $  -1.750 $ &   166.7  \\
 7751.12 & Fe~I  &    4.99  & $  -0.850 $ &    84.3  \\
 7780.57 & Fe~I  &    4.47  & $  -0.040 $ &   186.1  \\
 7807.92 & Fe~I  &    4.99  & $  -0.620 $ &    99.2  \\
 7941.08 & Fe~I  &    3.27  & $  -2.290 $ &    69.7  \\
 8239.13 & Fe~I  &    2.42  & $  -3.180 $ &    83.5  \\
 8293.49 & Fe~I  &    3.30  & $  -2.180 $ &    91.2  \\
 5197.58 & Fe~II &    3.23  & $  -2.230 $ &    78.6  \\
 5234.63 & Fe~II &    3.22  & $  -2.220 $ &    85.1  \\
 5991.38 & Fe~II &    3.15  & $  -3.570 $ &    38.3  \\
 6084.11 & Fe~II &    3.20  & $  -3.800 $ &    25.5  \\
 6149.26 & Fe~II &    3.89  & $  -2.690 $ &    38.9  \\
 6247.56 & Fe~II &    3.89  & $  -2.360 $ &    48.3  \\
 6369.46 & Fe~II &    2.89  & $  -4.200 $ &    22.4  \\
 6416.92 & Fe~II &    3.89  & $  -2.690 $ &    46.2  \\
 6432.68 & Fe~II &    2.89  & $  -3.740 $ &    36.5  \\
 6456.39 & Fe~II &    3.90  & $  -2.310 $ &    55.6  \\
 6516.08 & Fe~II &    2.89  & $  -3.450 $ &    54.2  \\
 5530.79 & Co~I  &    1.71  & $  -2.060 $ &    74.0  \\
 5647.23 & Co~I  &    2.28  & $  -1.560 $ &    51.5  \\
 6189.00 & Co~I  &    1.71  & $  -2.450 $ &    47.1  \\
 6632.45 & Co~I  &    2.28  & $  -2.000 $ &    42.2  \\
 7417.41 & Co~I  &    2.04  & $  -2.070 $ &    49.4  \\
 6053.69 & Ni~I  &    4.23  & $  -1.070 $ &    41.9  \\
 6086.28 & Ni~I  &    4.26  & $  -0.515 $ &    79.1  \\
 6128.97 & Ni~I  &    1.68  & $  -3.330 $ &    61.8  \\
 6130.13 & Ni~I  &    4.26  & $  -0.959 $ &    48.4  \\
 6175.37 & Ni~I  &    4.09  & $  -0.535 $ &    77.5  \\
 6176.81 & Ni~I  &    4.09  & $  -0.529 $ &    96.5  \\
 6177.24 & Ni~I  &    1.83  & $  -3.510 $ &    41.7  \\
 6186.71 & Ni~I  &    4.10  & $  -0.965 $ &    57.0  \\
 6204.60 & Ni~I  &    4.09  & $  -1.140 $ &    58.9  \\
 6360.82 & Ni~I  &    4.17  & $  -1.150 $ &    39.6  \\
 6370.35 & Ni~I  &    3.54  & $  -1.940 $ &    41.1  \\
 6378.25 & Ni~I  &    4.15  & $  -0.899 $ &    60.6  \\
 6482.80 & Ni~I  &    1.93  & $  -2.630 $ &    86.9  \\
 6586.31 & Ni~I  &    1.95  & $  -2.810 $ &    85.7  \\
 6598.60 & Ni~I  &    4.23  & $  -0.978 $ &    54.4  \\
 6635.12 & Ni~I  &    4.42  & $  -0.828 $ &    53.6  \\
 6643.63 & Ni~I  &    1.68  & $  -2.300 $ &   143.2  \\
 6767.77 & Ni~I  &    1.83  & $  -2.170 $ &   121.1  \\
 6772.31 & Ni~I  &    3.66  & $  -0.987 $ &    85.0  \\
 6842.04 & Ni~I  &    3.66  & $  -1.470 $ &    50.1  \\
 7422.27 & Ni~I  &    3.63  & $  -0.129 $ &   153.6  \\
 7797.59 & Ni~I  &    3.90  & $  -0.180 $ &   121.7  \\
 5782.12 & Cu~I  &    1.64  & $  -1.780 $ &   142.6  \\
 6362.34 & Zn~I  &   5.79 &   0.140 &    29.3  \\
 7800.29 & Rb~I  &   0.00 & 0.13 & syn \\
 6127.44 & Zr~I  &    0.15  & $  -1.06 $ &    18.0  \\
 6134.55 & Zr~I  &    0.00  & $  -1.28 $ &    16.2  \\
 6143.20 & Zr~I  &    0.07  & $  -1.10 $ &    20.4  \\
 5853.70 & Ba~II &    0.60  & $  -1.01 $ &    66.3  \\
 6141.70 & Ba~II &    0.70  & $  -0.07 $ &   130.0  \\
 6496.90 & Ba~II &    0.60  & $  -0.38 $ &   108.4  \\
 6390.48 & La~II &    0.32  & $  -1.41 $ &    12.6  \\
 6774.26 & La~II &    0.13  & $  -1.72 $ &    10.8  \\
 6645.11 & Eu~II &   1.38 &   0.12 &    15.4  \\
\enddata
\tablenotetext{a}{An inverted solar analysis was used
to determine $gf$.}
\end{deluxetable}

\begin{deluxetable}{lllr r | rr | c}
\tabletypesize{\footnotesize}
\tablenum{2}
\tablewidth{0pt}
\tablecaption{Abundances in \mystar
\label{table_abunds}}
\tablehead{
\colhead{Species} & \colhead{log[$\epsilon(X)$]\tablenotemark{a}} & 
\colhead{$\sigma_{obs}$\tablenotemark{b}} & \colhead{Num. of} &
\colhead{log[$\epsilon(X)/\epsilon(X)_{\odot}]$} &
\colhead{[X/Fe]\tablenotemark{k}} &
\colhead{$\sigma_{pred}$~for} &
 \colhead{Notes}  \\
\colhead{} & \colhead{(dex)} &  \colhead{(dex)} & \colhead{Lines} &
   \colhead{(dex)} & \colhead{(dex)} & \colhead{[X/Fe] (dex)} 
 }
\startdata 
Li~I & $\leq$0.86  & \nodata & 1   &  $\leq$+0.06 &  $\leq-$0.45 &    0.13 & syn   \\
C~I & 8.92   & 0.07 & 3 &  +0.31 & $-$0.12 &    0.21 & high $\chi$  \\
C(CH) & 9.05 & 0.15 & band  & +0.39 & $-$0.12 & 0.17  & syn  \\
O~I & 9.07   &  0.12 & 3 & +0.14 & $-$0.29 &  0.19 & high $\chi$ \\
Na~I & 6.64  &  0.11 & 4 & +0.77 &     +0.26 &    0.09 \\
Mg~I & 8.16  &  0.10 & 4 & +0.59 &     +0.08 &    0.07 \\ 
Al~I & 6.82 &  0.04 & 2 & +0.49 &     +0.13 &    0.08 \\
Si~I & 8.00 &  0.12 & 6 & +0.47 &     +0.04 &    0.17 & high $\chi$ \\
S~I & 7.62 & \nodata & 1\tablenotemark{h} & +0.45 &
        +0.02 &    0.20 & syn, high $\chi$ \\
K~I & 5.61  & \nodata & 1 & +0.38 &   $-$0.13 &    0.12 & c\\
Ca~I & 6.58  & 0.14 & 11 & +0.55 &     +0.04 &   0.07    \\
Sc~II & 3.71 &  0.03 & 2 & +0.54 &     +0.13 &    0.10 & d \\
Ti~I & 5.36  &  0.12 & 8  & +0.53  &  +0.02 &   0.11 \\
Ti~II & 5.57 &  \nodata & 1 & +0.67 &      +0.24 & 0.10  \\
V~I   & 4.47 &  0.06 & 5 & +0.68 &     +0.17 &    0.14 & d \\
Cr~I & 6.19 &  0.11 & 7 & +0.53 &     +0.02 &    0.07 \\
Mn~I & 6.03 &  0.11 & 3 & +0.66 &     +0.15 &    0.11 & e \\
Fe~I &  7.97 &  0.15 & 135 & +0.51 &  0.00  & 0.09\tablenotemark{i}  \\
Fe~II & 7.92 &  0.10 &  11 & +0.43   & $-$0.08 & 0.17\tablenotemark{j} \\
Co~I  & 5.55 &  0.14 &  5 & +0.77 &     +0.26 &    0.08 & d \\
Ni~I  & 6.86 &  0.13 &  22  & +0.67 &     +0.16 &    0.05 \\
Cu~I & 4.74 &  \nodata & 1 & +0.79 &     +0.28 &    0.15 & f \\
Zn~I & 4.97 &  \nodata & 1 & +0.42 &  $-$0.09 &    0.13 \\
Rb~I & 3.35 &  0.15 & 1 & +0.65 &     +0.13 & 0.17 & syn \\
Zr~I & 3.28 &  0.01 & 3 & +0.39 &   $-$0.14 &    0.16 \\
Ba~II & 2.50 &  0.05 & 3 & +0.38 &  $-$0.05 &    0.17 & d \\
La~II & 1.98 &  0.01 & 2 & +0.97 &     +0.54 &    0.12 & g \\
Eu~II & 1.38 &  \nodata & 1 & +0.88 &     +0.45 &    0.10 & g \\
\enddata
\tablenotetext{a}{This is log[$(n(X)/n(H)$] + 12.0~dex.}
\tablenotetext{b}{Rms dispersion about the mean abundance, using 
differential line-by-line abundances with respect to the Sun.}
\tablenotetext{c}{A 0.08~dex non-LTE correction relative to the
Sun is included for K~I.}
\tablenotetext{d}{The HFS corrections are small and not an issue.}
\tablenotetext{e}{The HFS corrections are large and are a concern.}
\tablenotetext{f}{The HFS corrections are very large and are a major concern.}
\tablenotetext{g}{Only one or two very weak lines detected.  Could be upper limits.}
\tablenotetext{h}{Very close pair of lines on wing of much stronger Si~I line.}
\tablenotetext{i}{The uncertainty in [Fe/H] inferred from the 135 Fe~I lines.}
\tablenotetext{j}{The uncertainty in [Fe/H] inferred from the 11 Fe~II lines.}
\tablenotetext{k}{The reference species (Fe~I or Fe~II) is given in
Table~\ref{table_sens_rel}.}
\end{deluxetable}

\begin{deluxetable}{l rrrr  }
\tabletypesize{\footnotesize}
\tablenum{3}
\tablewidth{0pt}
\tablecaption{Sensitivity of Deduced Absolute Abundances
\label{table_sens_abs}}
\tablehead{
\colhead{Species} & \colhead{$\Delta$log[$\epsilon(x)$] for}  
& \colhead{$\Delta$log[$\epsilon(x)$] for}
& \colhead{$\Delta$log[$\epsilon(x)$] for}
& \colhead{$\Delta$log[$\epsilon(x)$] for} \\ 
\colhead{} & \colhead{$\Delta$\teff\  +250~K} 
& \colhead{$\Delta$\grav\  +0.5~dex}
& \colhead{$\Delta v_t = +0.2$~\kms}
& \colhead{$\Delta$ [Fe/H] model +0.5~dex} \\
\colhead{} & \colhead{(dex)} & \colhead{(dex)}
& \colhead{(dex)} & \colhead{(dex)} 
 }
\startdata
Li~I  &         0.27 &        $-$0.02 &         0.00 &         0.01   \\    
O~I\tablenotemark{a} 
      &        $-$0.01 &         0.25 &         0.00 &         0.22   \\    
O~I\tablenotemark{b}  
      &        $-$0.32 &         0.17 &        $-$0.01 &         0.03   \\    
C~I   &        $-$0.29 &         0.17 &        $-$0.01 &         0.00   \\  
CH    &       0.20 &        $-$0.11 &        $-$0.05 &         0.04   \\
Na~I  &         0.18 &        $-$0.19 &        $-$0.02 &         0.12   \\    
Mg~I  &         0.08 &        $-$0.09 &        $-$0.03 &         0.06   \\    
Al~I  &         0.14 &        $-$0.06 &        $-$0.03 &         0.04   \\    
Si~I  &        $-$0.07 &      $-$0.01 &        $-$0.02 &         0.10   \\    
S~I   &        $-$0.21 &         0.16 &        $-$0.01 &         0.01   \\    
K~I   &         0.25 &        $-$0.24 &        $-$0.03 &         0.15   \\    
Ca~I  &         0.20 &        $-$0.11 &        $-$0.05 &         0.07   \\    
Sc~II &        $-$0.02 &         0.22 &        $-$0.04 &         0.18   \\    
Ti~I  &         0.25 &        $-$0.01 &        $-$0.02 &         0.02   \\    
Ti~II &        $-$0.05 &         0.22 &        $-$0.02 &         0.17   \\    
V~I   &         0.33 &        $-$0.03 &        $-$0.07 &        $-$0.01   \\    
Cr~I  &         0.18 &        $-$0.08 &        $-$0.04 &         0.06   \\    
Mn~I  &         0.20 &        $-$0.18 &        $-$0.07 &         0.16   \\    
Fe~I  &         0.12 &        $-$0.04 &        $-$0.05 &         0.10   \\
Fe~II &      $-$0.19 &           0.24 &        $-$0.05 &         0.19   \\
Ni~I  &         0.06 &           0.00 &        $-$0.05 &         0.13   \\
Co~I  &         0.12 &         0.05 &        $-$0.05 &         0.07   \\    
Cu~I  &         0.17 &        $-$0.11 &        $-$0.07 &         0.24   \\    
Zn~I  &        $-$0.13 &         0.13 &        $-$0.03 &         0.12   \\    
Rb~I  &         0.22 &        $-$0.03 &        $-$0.03 &         0.00   \\    
Zr~I  &         0.35 &        $-$0.02 &        $-$0.01 &         0.00   \\
Ba~II &         0.06 &         0.09 &        $-$0.11 &         0.26   \\    
La~II &         0.04 &         0.22 &        $-$0.01 &         0.20    \\   
Eu~II &        $-$0.02 &         0.22 &        $-$0.02 &         0.18   \\
\enddata
\tablenotetext{a}{for 6300~\AA\ line of [OI], only an upper limit here.}
\tablenotetext{b}{for the three lines of the 7770~\AA\ O~I triplet.}
\end{deluxetable}

\begin{deluxetable}{l rrrr r }
\tabletypesize{\footnotesize}
\tablenum{4}
\tablewidth{0pt}
\tablecaption{Sensitivity of Deduced Relative Abundances
\label{table_sens_rel}}
\tablehead{
\colhead{Species} & \colhead{$\Delta$[X/Fe] for}  
& \colhead{$\Delta$[X/Fe] for}
& \colhead{$\Delta$[X/Fe] for}
& \colhead{$\Delta$[X/Fe] for} 
& \colhead{Ref.\tablenotemark{a}} \\
\colhead{} & \colhead{$\Delta$\teff\  +250~K} 
& \colhead{$\Delta$\grav\  +0.5~dex}
& \colhead{$\Delta v_t = +0.2$~\kms}
& \colhead{$\Delta$ [Fe/H] model +0.5~dex} \\
\colhead{} & \colhead{(dex)} & \colhead{(dex)}
& \colhead{(dex)} & \colhead{(dex)} &  \colhead{}
 }
\startdata
Li~I  &    0.15 &    0.02 &    0.05 &   $-$0.09  & 1  \\    
O~I\tablenotemark{b} &   $-$0.13 &    0.29 &    0.05 &    0.12   & 1   \\    
O~I\tablenotemark{c}   &   $-$0.13 &   $-$0.07 &    0.04 &   $-$0.16 & 2   \\    
C~I   &   $-$0.10 &   $-$0.07 &    0.04 &   $-$0.19 & 2  \\   
CH    &     0.08 &   $-$0.07 &    0.00 &   $-$0.06   & 1   \\
Na~I  &    0.06 &   $-$0.15 &    0.03 &    0.02    & 1  \\    
Mg~I  &   $-$0.04 &   $-$0.05 &    0.02 &   $-$0.04  & 1    \\    
Al~I  &    0.02 &   $-$0.02 &    0.02 &   $-$0.06    & 1  \\    
Si~I  &    0.12 &   $-$0.25 &    0.03 &   $-$0.09   & 2 \\    
S~I   &   $-$0.02 &   $-$0.08 &    0.04 &   $-$0.18  & 2  \\    
K~I   &    0.13 &   $-$0.20 &    0.02 &    0.05     & 1 \\    
Ca~I  &    0.08 &   $-$0.07 &    0.00 &   $-$0.03   & 1   \\    
Sc~II &    0.17 &   $-$0.02 &    0.01 &   $-$0.01  & 2  \\    
Ti~I  &    0.13 &    0.03 &    0.03 &   $-$0.08    & 1  \\    
Ti~II &    0.14 &   $-$0.02 &    0.03 &   $-$0.02  & 2  \\    
V~I   &    0.21 &    0.01 &   $-$0.02 &   $-$0.11   & 1   \\    
Cr~I  &    0.06 &   $-$0.04 &    0.01 &   $-$0.04    & 1  \\    
Mn~I  &    0.08 &   $-$0.14 &   $-$0.02 &    0.06    & 1  \\    
Co~I  &    0.00 &    0.09 &    0.00 &   $-$0.03    & 1  \\    
Ni~I  &   $-$0.06 &    0.04 &    0.00 &    0.03   & 1   \\    
Cu~I  &    0.05 &   $-$0.07 &   $-$0.02 &    0.14   & 1   \\    
Zn~I  &   $-$0.25 &    0.17 &    0.02 &    0.02   & 1   \\    
Rb~I  &    0.10 &    0.01 &    0.02 &   $-$0.10   & 1   \\    
Ba~II &    0.25 &   $-$0.15 &   $-$0.06 &    0.07  & 2  \\    
Zr~I  &    0.23 &    0.02 &    0.04 &   $-$0.10   & 1   \\    
La~II &    0.23 &   $-$0.02 &    0.04 &    0.01   & 2 \\    
Eu~II &    0.17 &   $-$0.02 &    0.03 &   $-$0.01  & 2  \\    
\enddata
\tablenotetext{a}{1 denotes a value of [X/Fe] where Fe~I is used
as the reference, while for 2,
Fe~II is used.}
\tablenotetext{b}{for 6300~\AA\ line of [OI], only an upper limit here.}
\tablenotetext{c}{for the three lines of the 7770~\AA\ O~I triplet.}
\end{deluxetable}

\begin{deluxetable}{ll rr}
\tablenum{5}
\tablewidth{0pt}
\tablecaption{Diffuse Interstellar Bands in \mystar
\label{table_dib}}
\tablehead{
\colhead{Wavelength} & \colhead{FWHM} & \colhead{\eqw} &
\colhead{Central Depth}  \\
\colhead{(\AA)} & \colhead{(\AA)} &
\colhead{(m\AA)} &   \colhead{(\%)}
 }
\startdata
5778\tablenotemark{a}  &   9 & 1300 & 0.14 \\
5778\tablenotemark{b}  &   1.8 & 330 & 0.19 \\
6008 & 3.9 & 400 & 0.08 \\
6282 & 4.3 & 2000 & 0.30 \\
6611 & 1.5 & 200 & 0.14 \\
\enddata
\tablenotetext{a}{wide component of blend}
\tablenotetext{b}{narrow component of blend}
\end{deluxetable}

\clearpage

%

\end{document}